# Virtualization of Tiny Embedded Systems with a robust real-time capable and extensible Stack Virtual Machine REXAVM supporting Material-integrated Intelligent Systems and Tiny Machine Learning

**Stefan Bosse[1,2], Sarah Bornemann[3,4], Björn Lüssem[3,4]**

[1]University of Bremen, Dept. Mathematics & Computer Science, Bremen, Germany
[2]University of Siegen, Dept. Mech. Engineering, Siegen, Germany
[3]University of Bremen, Dept. Electrical Engineering, Bremen, Germany
[4]Institute for Microsensors, -actuators and -systems, Bremen, Germany

*Abstract* — **In the past decades, there has been a significant increase in sensor density and sensor deployment, driven by a significant miniaturization and decrease in size down to the chip level, addressing ubiquitous computing, edge computing, as well as distributed sensor networks. Material-integrated and intelligent systems (MIIS) provide the next integration and application level, but they create new challenges and introduce hard constraints (resources, energy supply, communication, resilience, and security). Commonly, low-resource systems are statically programmed processors with application-specific software or application-specific hardware (FPGA). This work demonstrates the need for and solution to virtualization in such low-resource and constrained systems towards resilient distributed sensor and cyber-physical networks using a unified low-resource, customizable, and real-time capable embedded and extensible stack virtual machine (REXAVM) that can be implemented and cooperate in both software and hardware. In a holistic architecture approach, the VM specifically addresses digital signal processing and tiny machine learning. The REXAVM is highly customizable through the use of VM program code generators at compile time and incremental code processing at run time. The VM uses an integrated, highly efficient just-in-time compiler to create Bytecode from text code. This paper shows and evaluates the suitability of the proposed VM architecture for operationally equivalent software and hardware (FPGA) implementations. Specific components supporting tiny ML and DSP using fixed-point arithmetic with respect to efficiency and accuracy are discussed. An extended use-case section demonstrates the usability of the introduced VM architecture for a broad range of applications.**

# Contents

**9. Conclusions**

**10. References**

# 1. Introduction

A significant increase in sensor density and sensor deployment, addressing ubiquitous computing, edge computing, as well as distributed sensor networks, requires advanced and robust data processing architectures. Fully digital sensors incorporating analog signal processing, digital signal processing, and communication are the state of the art (so-called "smart sensors"). Additionally, post-processing systems like data-driven Machine Learning (ML) are embedded in sensor nodes, too, opening the field of tiny ML for low-resource data-driven modeling. However, even small embedded microcomputers with chip areas less than 100mm2 provide low-power data processing (100 mW) with at least 100 MIPS computing power and several megabytes of memory. There is currently a high diversity of microcontroller architectures and devices. Prominent examples are low-priced ARM Cortex chips like the widely used STM32 series, or the ESP32 dual-core processor from a Chinese start-up company. There are different sensor and embedded system integration technologies: discrete application (e.g., to surfaces), surface integration, and most challenging, material integration. Material-integrated embedded systems are characterised by hard constraints, i.e., limitations in power supply, communication, and most importantly, inaccessibility after integration, preventing service and maintenance, including software updates. The exploration of the design space is a challenge.

This work focuses on **material-integrated embedded sensor systems**, used, e.g., for damage diagnostics by processing time-resolved ultrasonic signals (originating from guided ultrasonic waves, GUW).

An embedded system performing **sensor processing** consists of the following components:

1. Sensors and, to a lesser extent, actuators

2. Analog electronics for sensor signal processing;

3. Analog-Digital and digital-to-analog converters;

4. Data and code memory (typically with a multi-level architecture);

5. Data processing elements (with integer or fixed-point, and sometimes floating-point arithmetic units);

6. Communication controllers and technologies (wired and wireless, e.g., Bluetooth, WiFi-WLAN, Lora, and so on);

7. Energy is stored or harvested in some systems, but it is always managed.

There are passive and active measurement methodologies. For instance, acoustic emission (AE) analysis is a passive methodology that uses an environmental stimulus without requiring actuators (only sensors). GUW analysis requires an active stimulus, e.g., a sine burst signal applied to a piezoelectric actuator coupled to a surface or integrated in a material that generates waves measured by another piezoelectric sensor. The state of the art in GUW measurements is multi-path measurements and time-of-flight analysis. In this case, a communicating network of sensor nodes is required, e.g., to provide clock synchronisation. There is an additional challenge with materially integrated sensor networks.

**Architecture**s of sensor nodes and their application programming interface (API) can be strongly diverse and heterogeneous in combination in sensor networks. Commonly, program-controlled microcontrollers are the central processing units. The desing of embedded system includes architectur as well as system levels ([1]).

Application-specific designs using Field Programmable Gate Arrays (FPGA) are in the minority. Therefore, **virtualization** is even desired on tiny low-resource systems to overcome the heterogeneity gap [2], commonly isolating different architectures and systems (manufacturers) from each other [3], even in the case of application-specific SoC logic designs with FPGAs [4,5]. Virtualization can improve security, too [6], e.g., by using secure interpreters (like smalltalk presented in [7]). With regards to material-integrated embedded systems, industrial standards are not helpful, and oversized and common system virtual machines cannot be used.

**Hardware virtualization** can improve security in terms of robustness and failures, and is best supported by secure compilers [8].With the aid of a compiler and virtual machine, they take a two-stage approach to solving security issues at run-time: first, during the application development phase, a secure compiler fixes security flaws in the source code; second, while the applications are running, the secure virtual machine monitors unusual behavior like buffer overflow attacks or the handling of untrusted input data. In this work, such kind of security is provided by specific features of the processing architecture and the compiler.

Autonomous sensor and Internet-of-Things (IoT) nodes are **self-powered** (in the realm of the IoT often life-time battery powered), using energy harvesting to supply the node circuits with electrical energy, partly supported by energy storage to brick sourceless periods. NFC is a prominent example, serving as a communication and energy transfer platform, used in this work, too. Energy and power consumption are hard constraints in embedded systems, which require **energy management, energy prediction, and energy-based real-time task scheduling** [9] to serve important tasks first, addressed in this work by profiling and simple (optionally preemptive) task priority-based scheduling, discussed in Sec. 6.2. Premature energy loss is a common failure in autonomous, self-powered embedded systems.

This paper introduces and analyizes a real-time capable and extensible application-specific stack VM (REXA-VM) with some unique and special features that can be implemented in tiny embedded systems with a microcontroller and as little as 8 KB data RAM and 16 kB code ROM. One of the benefits of a virtualization layer is the freedom of implementation technologies. Therefore, an alternative implementation of the REXA-VM in FPGAs (or ASICs) with an RTL architecture using the ConPro High-Level Synthesis (HLS) [10] software is introduced and discussed, too. ConPro, unlike other HLS tools, models concurrent systems as a collection of concurrently communicating sequential processes, making it well suited for REXA-VM hardware implementation (in this case, a real virtualizing hardcore or softcore processor).The VM architecture is loosely based on earlier work [11], initially introducing a VM with code morphing capabilities as a core processor for mobile agents and supporting operationally equivalent software and hardware implementations. Stack processors processes commonly zero-operand instruction or zero data words.

The main application is the virtualization of autonomous (material-integrated) sensor and edge computing nodes with abstraction of sensors, actuators, user and device interfaces, communication hardware, and media-access and transport-layer protocols. The node's programming (Application Programming Interface, API) should always be pure textual, with no binary code provided from outside, i.e., the nodes should implement interpreters using just-in-time (JIT) compilers and a Bytecode VM.External communication with the VM is performed uniquely only by exchanging code frames with embedded data (active messages). Near field communication (NFC) is one target communication technology supported by the nodes and the VM to establish inter-node interactivity, introducing hard constraints due to communication-energy supply coupling. Inter-node application messaging is supported by simple data streams.

The instruction set architecture (ISA) of the VM can be freely defined and is fully customizable due to the deployment of VM program code generators. The VM consists of static (but parametrizable) source code and dynamic code generated based on specifications, e.g., the instruction word list. The code generators themselves can be modified and extended, too. There are code generators for the programming languages C and JavaScript, and the hardware description language VHDL (via HLS using the ConPro compiler [10]). Since REXA-VM is a stack processor, the Forth programming language [FORTH] is used as an initial set of instructions and programming model semantics.

Function calls are expensive, but stack machines have the advantage that operands are already on the stack, and function calls do not create additional overhead. Additionally, stack machines can process programs without any managed heap memory. Forth is a high-level stack machine programming language that has been widely used and well known for decades. Fourth instructions have a low variance in execution time, which is important for real-time systems. The instructions are commonly directly processed by a flat VM decoder loop.

The following sections discuss the constraints and requirements that must be satisfied, especially in but not limited to material-integrated sensor networks. Relevant aspects of software and hardware architecture, or virtualization layers and machines, are introduced. The REXA-VM is discussed in detail, and the relevant extensions (DSP, ANN/ML, Compiler) of the Forth-inspired programming language are defined and explained with examples. Design patterns are used to flesh out the internals of software and hardware implementations.

The **novelties** are:

1. Unified and customizable software and hardware implementations of the VM (H/W co-design at the architecture level), including operational equivalent simulators;

2. Highly customizable machine Instruction Set Architecture (ISA);

3. Just-in-time text-to-bytecode compiler (always bound to VM);

4. Security by design, i.e., tasks can be isolated without overhead or interference;

5. Portability and deployment in highly heterogeneous environments using various VM modes: VM can be used as a co-processor, embedded in any host application, or used independently to implement a complete sensor or computational node.

6. Low-resource requirements, optimized for ressource sharing, e.g., using an ADC sample buffer for computations from VM programming level, fast compared with other VMs; no data heap with memory management required;

7. Real-time capable VM due to multi-tasking, predictable instruction processing times, and micro-slicing with seamless integration in higher-level IO service loops;

8. An extensible and application-specific VM architecture is created by user-customizable and parametrizable VM code generation from code snippets and JSON configuration files, introducing a complete Hardware-Software-Simulation (HSS) co-design framework that supports the exploration of the design space under hard constraints.

The main advantage of the proposed VM architecture is the capability to create the main and crucial parts of the VM using code generators and adapt the VM architecture to specific applications and host architectures.

**Resilience and robustness on VM-level** is addressed by:

1. KISS: VM architecture is simple and provides inherent safety due to its simplicity;

2. Enhanced error detection and error recovery due to virtualization and isolation of critical architecture components; a pure textual code and data VM input interface increases the probability of detecting communication errors (data corruption);

3. Strict separation of control and data stacks (r-stack is not accessible by user code);2. Tasks can only access private data directly (data is embedded in their private code frames);

4. Ensemble VM execution (hardware or multi-core software implementation), executing the same code in parallel on multiple VM instances and comparing intermediate states and results  (majority decision making; stopping of faulty computations);

5. Check-pointing with optional persistent storage enabling stop-and-go (instead of stop-and-forget) processing (e.g., on irregular and short power cycles);

6. Exception handling;

7. Adaptivity due to incremental code execution (i.e., code updates overwriting older code via the global dictionary);

8. Hardware-Software-Simulation Co-design by unified DB-driven VM code generators enables the operational simulation, profiling, and test of real network nodes with the same operational semantics and discrete timing;

9. Optional special data codings (hardware, simulated in software) for improved error detection and error correction.

The conceptual overview of the DB- and code generator-driven REXA-VM development flow is shown in Fig. 1.

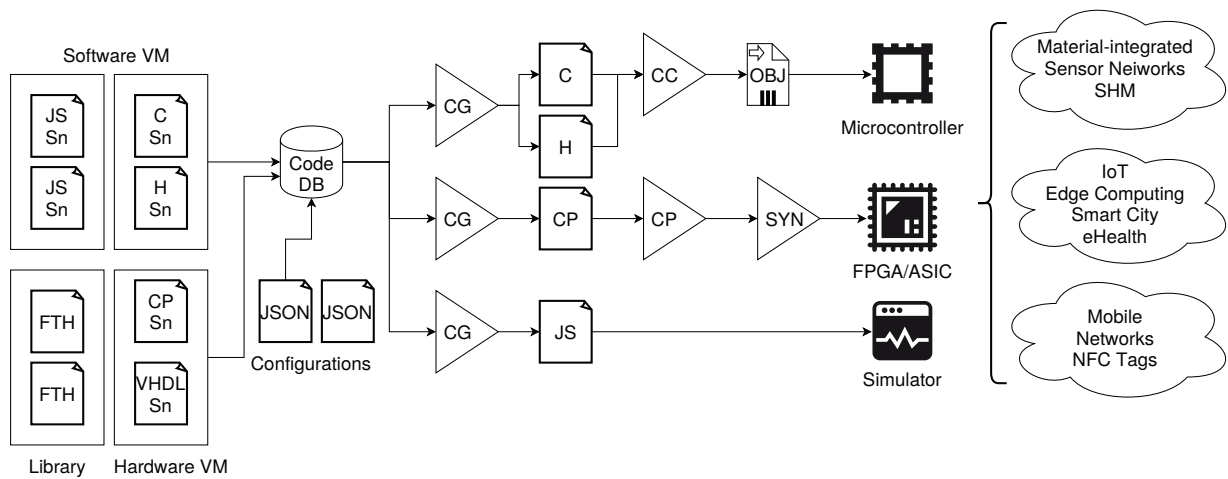

**Fig. 1.** Overview of the overall concept of REXA-VM development (C-SN: C source code snippet, H-SN: C Header snippet, JS: JavaScript, FTH: Forth VM code definitions, JSON: JavaScript Object Notation, CG: Code generator, CC: C Compiler, CP: ConPro HLS, SYN: RTL synthesis tool)

The proposed VM architecture and design methodology (using software and hardware implementations) address the following **application fields**:

1. Material-integrated sensor networks and intelligent systems in general;

2. Structural Health Monitoring

3. Highly miniaturized IoT devices and edge computing networks;

4. Smart City, eHealth

5. Mobile networks and active and adaptive NFC tags.

## 2. Material-integrated Sensor Nodes

### 2.1 Taxonomy and Efficiency

There are special constraints, the most important of which are:

- Silicon circuits typically require chip die thinning technologies due to their small size and resistance to mechanical stress, bending, and stretching.

- No (or limited) service or maintenance is available.

- Low power, energy, and resource consumption (8 kB RAM, = 32 kB ROM, 1-20 MHz clock, 1-20 MIPS native processor speed)

- Self-powered nodes: energy harvesting is not deterministic (longitudinally and vertically);

- No continuous operation; interrupted by operational sleep modes; non-deterministic due to power outages and interruptions; + Increased failure rates (partial, data, communication, and program code corruptions; processing and communication interruptions; communication loss; total node failure);

Efficiency of data processing is always an important objective to optimize, especially for material-integrated sensor networks. The efficiency of data processing systems can be compared by the following normalized performance factor $\epsilon$:

$$\epsilon = \frac{C \cdot M}{A \cdot P} \tag{1}$$

with $C$ representing a data processing system's computational power in instructions per second (MIPS), $M$ representing memory capacity (RAM/ROM) in k Bytes, $A$ representing the entire chip area in $mm^2$, and $P$ representing electrical power consumption in mW. This is indeed a very rough efficiency factor computation, but it enables the comparison of different architectures and processors. Note that the chip area is commonly neglected, but it is a key factor for material integration. The retrieval of the pure chip die area is difficult due to missing information. Therefore, x-ray diagnostics were used to measure the chip die area. Any sensors and electronics embedded in materials have a physical and mechanical impact on the host material, weakening the physical and mechanical features of the material. Silicon-based chips can be bent but not stretched. By reducing the chip area, you reduce the risk of mechanical failure and material impact.

A summary of parameters and efficiencies for common microcontrollers suitable for tiny embedded systems is shown in Tab. 1, compared with some FPGA devices. The selection of a suitable device depends on the complexity and time constraints of the data processing, communication, size, and energy supply. In this work, a particular sensor node with a STM32 ARM Cortex M0+ (L031G6U6) microcontroller is used. The sensor node receives energy from an NFC radio antenna (see use-cases, Sec. 7.1). The harvester can collect and deliver up to 15 mW of continuous power (as long as the electromagnetic field is applied), significantly narrowing the set of suitable devices. Additionally, energy management at the task-scheduling level is required. For instance, if the processor performs computations, no external communication is possible in parallel!

Besides power consumption and energy management, robustness against failures and active resilience are important key features, including mechanical and structural properties. Resilience is commonly addressed at the middleware (software) and system levels, e.g., by applying agent-based methods (ABM/ABC). Resilience defines basically a gap between required and delivered capabilities over time [12], e.g., redundancy, adaptivity, adaptive fault tolerance, and most importantly, error detection on various levels (hardware and software, sensors, processors, and so on).

The computational efficiency with respect to power consumption, chip die area, and memory differs highly among different microcontrollers. From the table the selection of the L031 device and L073 in future versions of the aforementioned sensor node can be clearly derived from ε.

| Device | Chip Area | Clock/MIPS | Power | RAM/ROM | ε |
|---|---|---|---|---|---|
| Atmel Tiny 20 | 2.1 mm$^2$ (1.55x1.4x0.53 mm) | 12 MHz | 4 mW | 0.1 kB/2 kB | 3 |
| ARM Cortex M0 (Smart Dust 2002) | 0.1 mm$^2$ | 740 kHz | 70 mW | 4 kB/4 kB | 0.84 |
| FreeSclae KL03 (ARM Cortex M0+) | 4 mm$^2$ | 48 MHz | 3 mW | 2 kB/40kB | 168 |
| STM32 F103C | ~5 mm$^2$ | 72 MHZ | 100 mW | 48 kB/256 kB | 44 |
| STM32 L031G6U6 | 0.25 mm$^2$ (meas.) | 16 MHZ | 2 mW | 8 kB/32 kB | 1280 |
| STM32 L073CZU6 M0+ | ~1 mm$^2$ | 16 MHZ | 3 mW | 20 kB/192 kB | 1130 |
| Xilinx Spartan 3-500E | 9.6 mm$^2$ (meas.) | 50 MHz | 100 mW | 45 kB | 2.34 |
| Xilinx Spartan 7-S25 | ~50 mm$^2$ | 100 MHz | 100 mW | 202 kB | 4 |

**Tab. 1.** Computational efficiency of common microcontroller-based embedded systems (∗ used in this work), compared with FPGAs. It is assumed: 1 Hz clock frequency = 1 native machine code IPS

## 2.2  Software Architectures

An example and the root of this work of a self-powered and autonomous wireless material-integrated sensor node is shown in Fig. 2, developed for data-driven damage diagnostics in Fibre-Metal laminate structures, e.g., used for aircraft technologies. Energy is supplied to the sensor node from outside sources via RFID/NFC. The current node version has no energy storage. Therefore, the sensor node faces hard power constraints. The energy harvester can deliver up to 15 mW of power continuously, limiting the design space significantly. The energy harvester has to supply the microcontroller, the analog signal processing, and the NFC communication controller. An ARM Cortex M0 with about 1-10 mW of power dissipation satisfies the power constraints while still providing enough computational power to perform DSP, ML, and processing the VM. Details about the sensor node can be found in use-case Sec. 7.

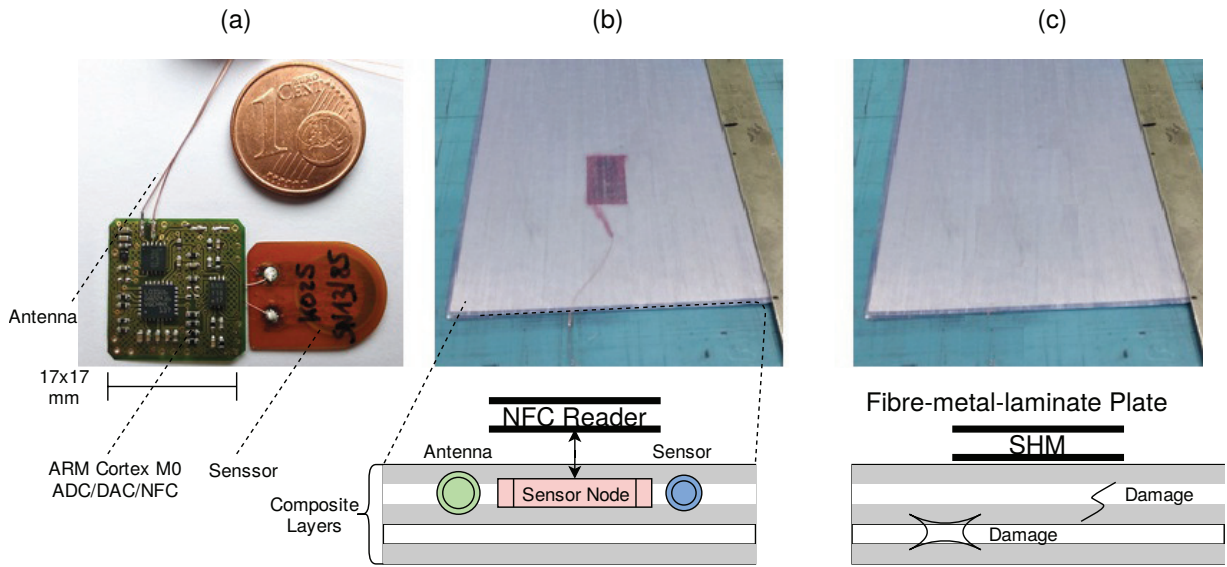

**Fig. 2.** Example of material integration of a fully equipped wireless sensor node including antenna and piezoelectric sensor in a FML plate used in this work (a) Sensor Node (b) Embedding of Sensor Node in Fibre layer (cut-out integration) (c) Final coverage with metal layer

## 2.3 Hardware Architectures

Besides commonly and widely used microcontroller architectures providing program-controlled data processing, application-specific hardware, e.g., using Field Programmable Gate Arrays (FPGA), is a suitable alternative for a data processing and communication platform, posing several advantages:

1. Application-specific operations;

2. Application-specific architecture optimizing energy consumption, size, and integration (one-chip solutions);

3. Parallel data processing on the control- and data-path levels;

4. Optimal efficiency ε with respect to chip area (transistor count), power consumption, and computational power.

Fig. 2 shows a prototype of a sensorial material equipped with a distributed sensor network (FPGA nodes with ADC), using strain-gauge sensors attached to the surface of a rubber plate. The job of the distributed sensor network is to detect externally applied loadings on the plate, providing a classification and localization of loads using ML.

The advancement of technology has resulted in low- and very-low power microcontrollers, which typically consume between 0.1 and 0.5 mW per MHz of clock frequency, typically scaling 1:1 with respect to instruction throughput performance (i.e., 1 MHz = 1 MIPS). On the one hand, hard-core implementation of application-specific circuits using ASIC technologies can always outperform generic microcontrollers (e.g., by utilizing clock gating methods). On the other hand, FPGAs are generic logic fabrics with high overhead due to their configurability, typically resulting in 10 mW / MHz power consumption. SRAM-based FPGA technologies require a configuration bitstream loaded each time into the FPGA to configure the logic matrix, requiring a high start-up power, not compatible with self-powered energy harvesting nodes. Flash-EEPROM-based FPGAs do not require initialization, but also show higher power consumption than very low-power microcontrollers, assuming a logic gate count of about 500k - 1M gates. To summarize, FPGAs are a valuable rapid prototyping technology but are not suited for very low-power large-scale circuits. We will consider FPGA-based implementations of the VM, too, but only to show the suitability of the proposed design approach and architecture, eventually creating low-power ASICs.

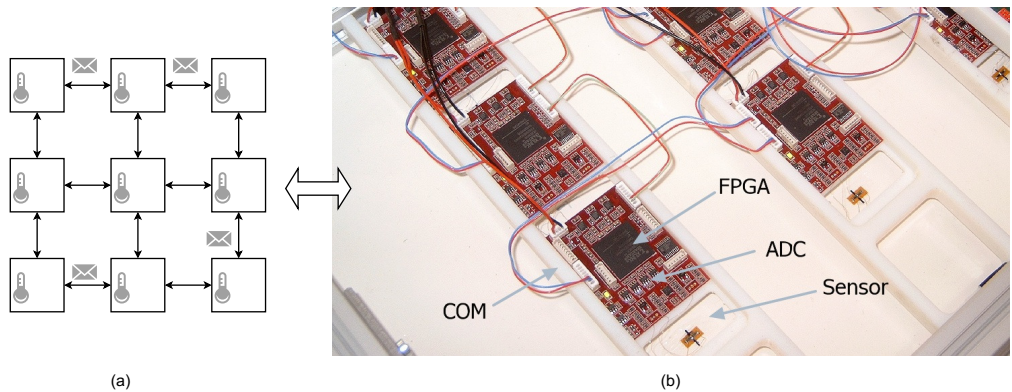

(a)                                        (b)

**Fig. 3.** Example of a sensor node applied to a material surface (rubber plate) with an application-specific processor using an FPGA (Xilinx Spartan 3-1000) and some external analog components with 300 kSPS ADC and DAC (a) Communication network topology (b) Back side of rubber plate

## 2.4 Sensors and Signal Processing

## 2.5 Communication

There are two classes of communication addressed in this work:

1. Communication between the VM and the host application, including sensors, actuators, and other peripheral devices;

2. Network communication among nodes of a sensor network.

Communication can be further classified in data-driven and event-based synchronisation.

### 2.5.1 Network Communication

Physical network communication among nodes can be classified in (see also Fig. 4):

1. Wired peer-to-peer communication (electrical or optical connections); composition of Transputer-like mesh networks (e.g., [13]) with simple high-speed serial-link communication;

2. Wireless peer-to-peer short-range communication (e.g., ZigBee or Bluetooth, RFID/NFC);

3. Wireless broadcast short-range communication (e.g., Bluetooth, NFC);

4. Wireless multi-cast mid-range communication (e.g., WiFi WLAN);

5. Wireless single- or multi-cast long-range communication (e.g., LoRa WAN or cellular mobile networks).

### 2.5.2 RFID/NFC Technologies

Wireless communication is used by material-integrated as well as distributed tiny sensor or communication nodes. RFID and NFC technologies are widely used communication standards that supply nodes with electrical power, too. The typical electrical power that is delivered by passive NFC tag controller circuits is about 10-20 mW, which is used for digital signal processing as well as for sensing and communication.

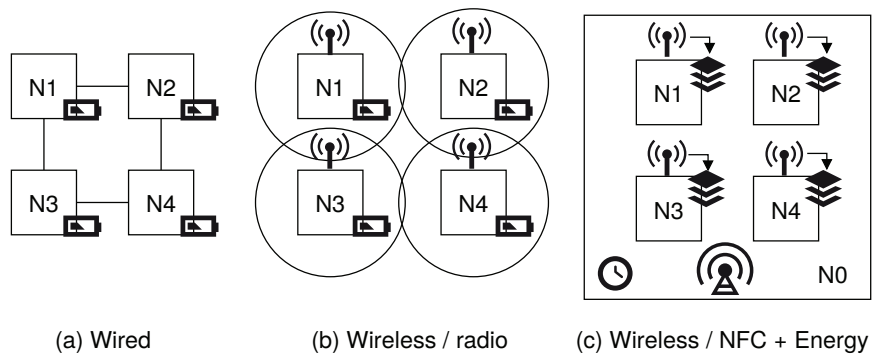

(a) Wired                (b) Wireless / radio                (c) Wireless / NFC + Energy

**Fig. 4.** Different communication architectures

Since one node cannot communicate directly with another, a repeater solution is chosen by the master NFC reader. Node group communication is required, for instance, for clock synchronisation (node 0 in Fig. 4).

## 2.6 Failure Taxonomy and Models

Material-integrated sensing systems operate under harsh environmental conditions. In principle, any failure can occur. The main question to be answered by an automated process is the detection and classification of sensor node failures.

In particular, there are:

- Sensor failures:

  - Increased noise

  - Decalibration

  - Permanently stuck fault

  - Sensor decoupling from material

  - Interference

  - Cracks and delaminations of sensor materials

- Energy supply interruptions:

  - Reduced or complete power loss (longer off-times)

  - Degraded or defect energy storage

  - Electronic faults in power supply circuits

- Increased noise and over- or under-voltage
- Communication failures:
  - Complete loss of connectivity
  - Partial loss of messages
  - Errors in data (bit flipping)
  - Connection faults in  (wired, electrically, optically)
- Fraud and side attacks
- Data processing errors:
  - Interrupted control flow
  - Invalid data and code
  - Program errors
  - Processor errors

The data processing system must be robust against data and power supply errors, e.g., by redundancy or improved error checking, and by architectural design, as discussed and addressed in this work. Noise or instability on power lines can cause data corruption (RAM and ROM). A faulty operation can result in interrupt traps, which must be handled to reset the device properly, optimally using check-pointing and recovery by state saving and restoring after re-boot.

## 2.7  Real-time Systems and Energy

Recently, energy harvesting has become a practical method for extending or establishing the operational time of sensor networks. Common power management techniques, however, must be rethought if every network node is powered by a variable energy source. This is true in particular if a certain application's real-time responsiveness needs to be ensured. The energy source's characteristics, the energy storage system's capacity, and the individual activities' due dates should all be taken into consideration when scheduling tasks at single nodes. In the context of regenerative energy, greedy scheduling algorithms (like the Earliest Deadline First EDF) are inappropriate [9]. Instead, as proposed in [9], a Lazy Scheduling Algorithm (LSA) is used.

One of the advantages of the REXA-VM presented in this work is the average constant run-time and, hence, energy consumption of VM instructions and the anytime interruptable VM instruction loop. Additionally, profiling of user-defined and externally provided IOS functions is supported, enabling predictive scheduling policies. REXA-VM can be implemented in software as well as in hardware, both relying on an energy-driven task and thread scheduling (threads are separate FSM processes that can be controlled by an energy and scheduling manager). There is an energy-driven scheduling scenario for a system whose energy storage is recharged by an environmental source, with significant support from the presented REXA VM. There are two measures, and their prediction is important: future power and energy consumption and future power and energy harvesting.

The algorithms and challenges of real-time scheduling of energy harvesting sensor nodes are discussed in Sec. 6.2.

## 3. Virtual Machine Architecture

The REXA VM is a highly configurable high-level stack processor that executes Bytecode generated from textual program code, ensuring high portability, flexibility, and robust code processing in heterogeneous environments (including the deplyoment of different versions of the REXA VM). Basically, the entire ISA can be customized, creating a balance between generality and application-specific optimizations. Most instructions are zero operand operations accessing data entirely on the VM stacks (in the program code model, they are post-fix instructions because data was stored by previous operations).

The following Fig. 5 shows the principle architecture of the REXA VM and its just-in-time (JIT) compiler. The architecture details depend on the configuration (single- or multi-tasking, number of stacks, and customized extensions and accelerators). The principle architecture is equal for software and hardware implementations. The native compiler is available in software and in hardware, but can be replaced by a lightweight pure Forth-based implementation. Profiling is an optional feature used for predictive real-time scheduling, as well as the energy-aware real-time scheduler.

The code segment (CS) is the central storage for source code, bytecode, and embedded data. The CS is partitioned into dynamically sized code frames, commonly assigned to a task (depending on the scheduling model). The scheduler controls and monitors the Bytecode loop (*vmloop*). Code operations can suspend task execution by waiting for events, handled by an event table. The Input-Output System (IOS, similar to the widely used Foreign Function Interface, extends the code and data space of the VM). The VM architecture is optimized for ressource sharing, e.g., using an ADC sample buffer for computations from VM programming level, too.

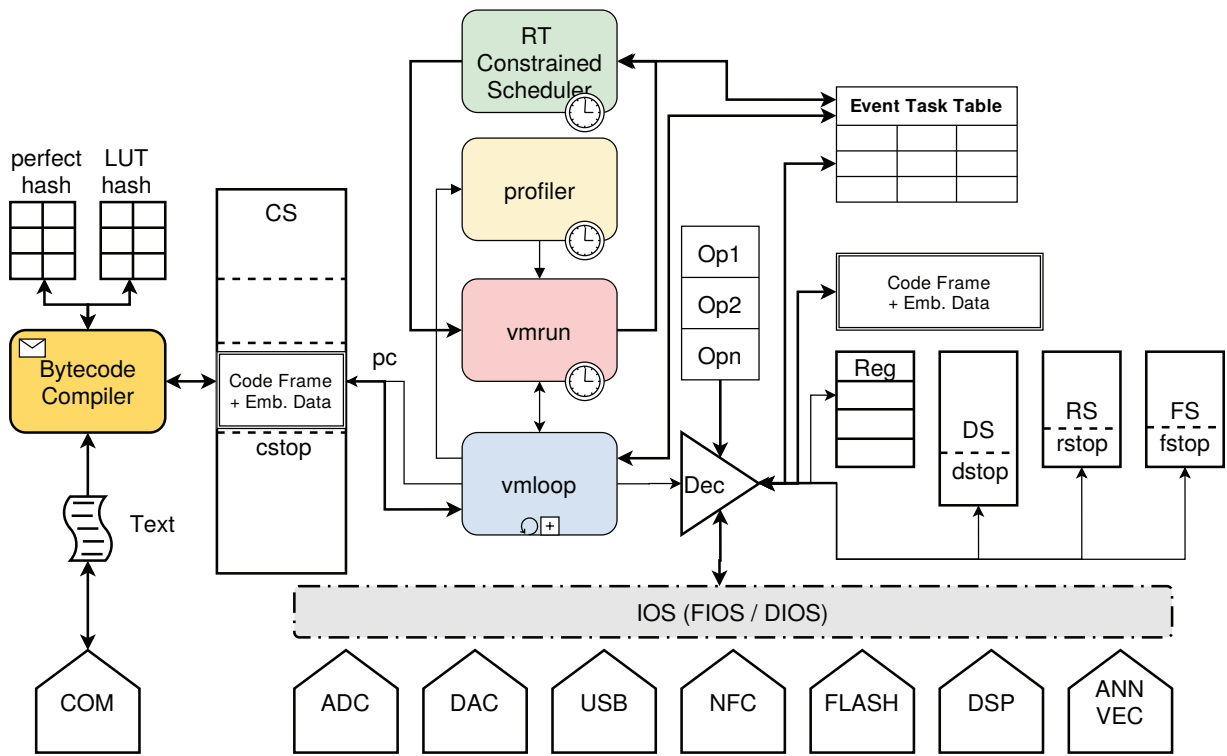

**Fig. 5.** Basic REXA-VM architecture with integrated JIT compiler, stacks, and Bytecode processor

### 3.1 Code Segment and Frames

The main user program memory is the code segment of the VM is the code segment, (CS). A code segment is organized in byte cells and has a static fixed size. The new program code allocates a part of the CS, called a code frame (CF). A code frame can contain top-level operations, word definitions that can be added to the VM dictionary, and data variable allocations embedded in the code frame (there is no heap memory), either directly between Forth instructions or at the end of the code frame program (e.g., non-initialized array data). After the code frame program processing terminates (called the *end* operation), the code frame with all of its data is removed. Alternatively, the code frame can be kept alive after termination, and exported code word references can be used from later code frames. This feature enables incremental and partitioned program execution.

In single-tasking mode, the code segment is commonly incremental and persistent. That means, program code is added to the CS incrementally. Each new code fragment is compiled and immediately executed. The current program terminates with the *end* instruction, with or without persistence. Without persistence, the code is removed after termination. If the code fragment is persistent, a new code fragment is stored at the end of the previous one. Persistent code cannot be removed, only resetting the CS is possible. Only the current code fragment can be scheduled, without branching to other code regions. In multi-tasking mode, the code segment is partitioned into dynamically sized code frames. Code frames can be removed later, and code frames can be linked, as illustrated in Fig. 6. Scheduling can branch to other code frames. Each code frame is associated with a task, except code frames that terminated with persistence.

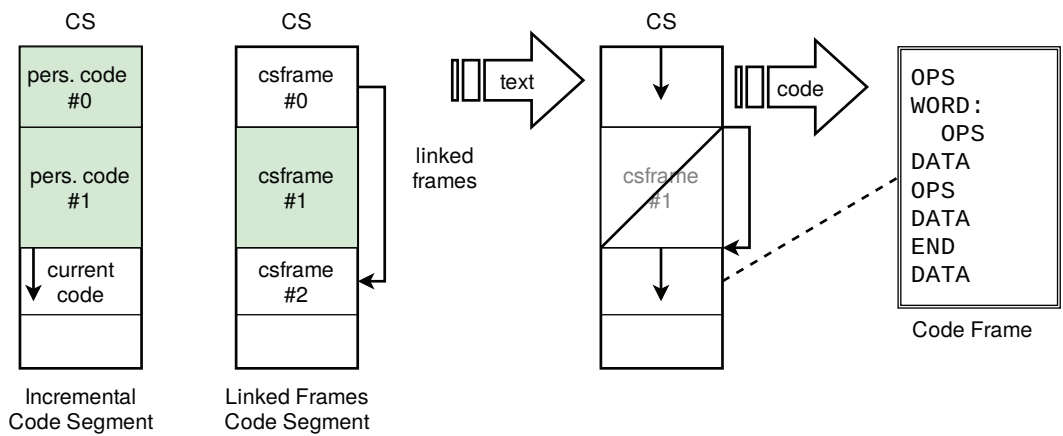

**Fig. 6.** (Left) Incremental growing code segment (single-tasking), persistent code cannot be removed (Right) Dynamically partitioned code segments using code frames and linking of code frames due to fragmentation

A code frame merges code and data. The data (scalar and array variables) can only be accessed by code inside the code frame. Operations can be bound to named words, which can be exported in the global dictionary and accessed from other code frames (and tasks). If a function word is exported, the code frame is locked and is not removed if the code processing reaches the end (instruction).

### 3.2 Stacks

Temporary (short lifetime) data is stored and manipulated directly on fixed-size stack memories:

1. The Data Stack (DS) holds most of the processing data and instruction operands;

2. The Return Stack (RS) for function calls (not accessible from the programming level for security reasons);

3. Optional a loop stack (FS) used for loop counters and secondary user data (can be merged with RS for memory efficiency).

All non-temporary data is either embedded in the code frames or provided by the host application via the Data Input-Output System layer (DIOS) API.

The stack cell width is always 16 bits (single word width). The REXA VM supports double word operations, too (as a configurable option). Double words are composed of two single data words (word order depends on the native byte order of the underlying processor). Double words can be directly read and written from and to stacks by the VM (single memory access). The access time of multiplexed single and double word access to the stacks in software by memory pointer casting is commonly identical (assuming 32-bit microprocessors). The hardware implementation can split the double word access into two memory cycles or use 32-bit memory for stacks, always providing one-cycle memory access. For performance reasons, the hardware implementation can implement stacks with block RAM components and single registers holding the first top values, enabling parallel computations on stack elements. The push and pop operations involved in most of the VM instruction code words modify stack pointers (*dstop*, *rstop*, *fstop*). For security reasons, the return stack (which holds code pointers on function calls) should not be accessed directly by program code.

Besides hardcore stacks implemented inside the VM, softcore stacks can be implemented on the programming level in data arrays (embedded in code frames). Push and pop operations are provided by the core instruction word set:

```
array mystack 100
1 mystack push
mystack pop . cr
3 mystack get ( Gets copy of n-th value from top )
```

### 3.3 Multi-tasking

The VM architecture and code processing can be either single-tasked or multi-tasked (using the code frame mode). Although the VM can be replicated and share the same code segment and globals, providing multi-threading (real multi-tasking) with parallel execution on multi-core processors or replicated RTL hardware, multi-tasking is a scheduling without concurrency, i.e., tasks are co-routines.

Common non-preemptive scheduling points are summarized in Def. 1. The *task* command explicitly creates a new task in multi-tasking mode. The `end` command indicates the end of code frame processing, leaving the VM execution loop. If there are no exported words or pending tasks, the code frame will be removed. If energy-aware real-time scheduling is enabled, a task can be optionally assigned a priority and a deadline. The priority can be changed by the scheduler only if there are energy conflicts. There are event-based and computation-based tasks. Both differ in their run-time behaviour. Negative priorities indicate a short-running event-based IO task; positive priorities indicate a greedy computational task.

**Def. 1.** Scheduling points in multi-tasked VM processing and task creation

```
yield
<millisec> sleep
<millisec> <value> <var> await
receive
```

```
inp
<priority> <deadline> $ <codeword> task  ( number number functionaddr -- taskid )
end
```

### 3.4  Parallel VM

Besides multi-tasking, which provides a scheduled programming and code execution model without parallelism, multiple VM instances can be easily composed into a parallel VM. Each parallel VM shares the same code interpreter (decoder) and code segment (and compiler, if implemented), but with individual stack segments and VM registers.

### 3.5  Synchronisation and Communication

There are two classes of synchronisation provided by REXA-VM, as summarized in Tab. 2:

1. Internal temporal and event synchronisation, e.g., waiting for a specific time interval or the completion of a measuring job (or both);

2. External inter-node synchronisation and communication with data exchange in a sensor network, e.g., for clock synchronisation or sensor data requests.

Using simple *receive* and *send* communication operations, as shown in Tab. 2, Transputer-like mesh-grid networks can be established with little overhead. The host application software or dedicated devices in a hardware SoC design implement the physical link layer and communication between nodes. The *receive* operation is a synchronous operation that blocks the current task (or the entire VM) until a new data token is received. The *receive* operation is ideally embedded in a separate receiver task.

| Operation | Operands | Description |
|---|---|---|
| sleep | ( millisec -- ) | Suspends task execution for a fixed time interval. |
| await | ( millisec value varaddr -- status ) | Suspends task execution for the arrival of an avent, given by the change of an atomic guarded variable to the specified *value* or a time-out occurs. |
| out | ( value -- ) | Outputs a value to a generic stream device (e.g., connected to a master controller or cimputer), commonly using a multiplex standard ouput channel. |
| in | ( -- value ) | Receives a data value from a generic stream device. |
| send | ( value dstaddr -- ) | Sends one data value to the specified node or communication link. |
| sendn | ( length offset dataaddr dstaddr-- ) | Sends *n* data values from an array buffer to the specified node or link. |
| receive | ( srcaddr -- value ) | Receives one data value from source node or link. |

**Tab. 2.** Event-based internal synchronisation operations and Data-driven and stream-based external synchronisation and inter-node communication, which suspends virtually the current program processing by leaving the current VM interpreter loop round.

### 3.6 Input-Output System (IOS)

The Input-Output System (IOS) is similar to the widely used Foreign Function Interface (FFI) and provides unified host application integration and extends the FORTH word set with bridged native C/C++ functions. Additionally, host application variables (scalar and numerical array types) can be directly accessed from VM programs. Most of the DSP and ML operations are not part of the core REXA-VM engine. Instead, they are added on demand by the host application from customizable libraries.

The basic IOS API (C) and some examples are shown in Def. 2. Data is passed to an external function via the data stack (*args* values are popped from the stack), and optional return data (one value) is pushed on the data stack. The data size of values can be in the range of 1 to 4 bytes. Values with a data size of 4 bytes (32 bits) are divided into two values (the most significant and least significant words). The word order is architecture-dependent.

---

**Def. 2.** IOS API enables the extension of the REXA-VM instruction word and variable set.

---

```
 1: u8 fiosAdd(char *name,
 2:             void (*callback)(),
 3:             u8 args,
 4:             u8 argsize,
 5:             u8 retsize);
 6: u8 diosAdd(char *name,
 7:             void *data,
 8:             u16 cells,
 9:             u8 size);
10: // Examples
11: // milli ( -- msw lsw )
12: u32 myMilli() { .. }
13: fiosAdd("milli",IOSCALLBACK(myMilli),0,0,4);
14: // <trigMode> <depth> <ampGain> <sampleFreq> adc
15: void myADC(u16 trigMode, u16 depthIndex,
16:             u16 ampGain, u16 sampleFreqKS) { .. }
17: fiosAdd("adc",IOSCALLBACK(myADC),4,2,0);
18: u16 SampleBuffer[8192];
19: diosAdd("sample",IOSDATA(SampleBuffer),8192,2);
```

---

### 3.7 System Call-gate Interface

The system call-gate interface is a unified communication and execution interface to the REXA VM run-time environment and compiler. There are two complementary versions of the system call-gate interface addressing software and hardware implementations of the VM:

1. A shared-memory architecture providing operational access to the VM and the compiler via one universal *vmsys* function. The host application runs or compiles code via protected *vmsys* calls (similar to the kernel system call interface of modern operating systems). Additionally, the host application can install callback handlers, e.g., for console output or for communication instruction words.

2. A message-based architecture providing operational access to the VM and the compiler via a communication channel and remote-procedure calls. This interface enables the integration of separated (hardware) REXA processors with existing microcontroller systems, e.g., via a bus system or a serial link. Mesh networks as well as wireless networks can be easily constructed.

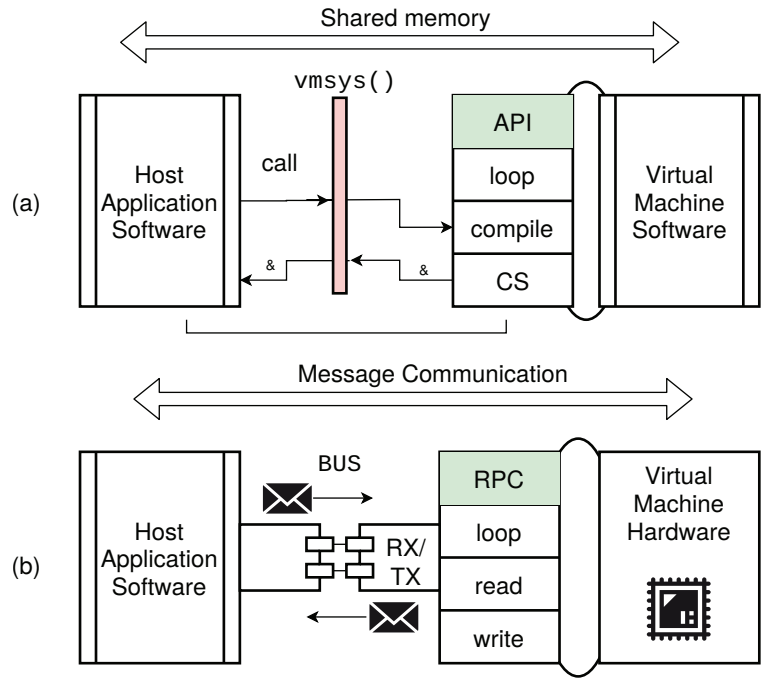

**Fig. 7.** System Call-gate Interface connecting a sensor node root application software to an isolated VM instance (a) via shared memory and a single system call function (b) via message-based communication and a serial link or signal bus

The software call-gate version supports direct sharing of data and function pointers from the VM to the host application and vice versa. For example, the VM code segment *CS* can be shared directly with the host application, or the host application can provide printer and communication functions that can be accessed directly from the VM and Forth words.

### 3.8 Exception Handling

One of the major features of robust and resilient data processing is exception handling. Exceptions can occur due to non-deterministic failures (hardware, software, ISA, user), or programmatically to leave the current control flow environment. Exceptions are handled by generic user-defined word functions, attached to an exception signal (or event), as shown in Def. 3. There are some predefined exceptions that can be bound to a word function. Examples are `trap`, `stack`, `interrupt`, `io`, `timeout`, or `divbyzero`.

The interrupt exception, e.g., is raised if the run-time slice of the current task or the entire VM is interrupted prematurely, e.g., due to a awaited lack of energy. There is no *try-catch* environment. Instead, catch points can be set. After an exception was handled, the VM branches to the last catch point. If there is a pending exception, the catch instruction pushes the exception onto the stack, otherwise, a zero value indicates no pending exception, which can be evaluated by a following conditional branch.

Branching to a catch point aligns the return stack with the state at which the catch point was installed.

---

**Def. 3.** Exception API enables handling of failures and errors.

---

```
: <handler> ... ;
( Bind <handler> to exception <exc> )
$ <handler> exception <exc>
catch if ... <exception occurred> ... endif
<exc> throw
```

---

### 3.9 Compiler

The compiler translates the source code text into bytecode instructions. It is a just-in-time (JIT) compiler that can compile code incrementally and on demand. Since the ISA of stack processors consists mostly of zero-operand instructions, it supports fine-grained compilation at the token level. The source text can be directly stored in the code segment referenced by a code frame (or any other data buffer, alternatively). Most instruction words can be directly mapped to a consecutively numbered operation code. Therefore, the compiler translates the source code into bytecode directly in-place, i.e., by replacing the text with bytecode, saving additional target memory buffers. An instruction word consists of at least one character, and thus can always be replaced by the op-code (one byte). Although a literal value can consist of only one digit and the data of a single word value occupies two bytes, there is always a space or newline character after a literal value, providing the required data space. Extension of the current code frame at the end is always possible (as long as there is free space in the CS). One exception is a double word literal value requiring at least two characters and the suffix "l", followed by an obligatory separator character and the space, providing four bytes of data space in total.

Data is either stored on the stacks during run-time or is embedded in the code frame during translation. Scalar variables and initialized arrays can always be embedded in-place. Non-initialized arrays are appended at the end of the compiled code frame.

The compiler is part of the VM and is processed directly on the (low-resource) embedded system or in hardware. This ensures security checks and guarantees that a task assigned to a code frame can only access its private data. Finally, the textual interface ensures compatibility and interoperability among different VM versions (with different binary ISA, but a common sub-set). Alternatively, most parts of the compiler can be implemented in the high-level VM language itself. In this case, the VM only provides basic compiler operations like a tokenizer and a dictionary.

To enable fast compilation (with respect to low-resource microcontrollers), no traditional lexer and parser are used. Instead, the parsing process uses two different approaches:

1. A perfect hash table [14,15] (PHT) for core words (more than 100 core words) with a constant search time. Because the hash index is directly linked to the operational code, the hash table itself must be saved. A hash function, on the other hand, cannot detect words that do not match any of the core word set.Therefore, a string table is required containing all core words indexed by the same hash index, finally comparing a hash-predicted word from this table with the current word to be parsed.

2. A Linear Search Table (LST) with non-constant search time. The LST implements an iterative and sequential character search in a compacted linear array, as discussed below. The LST needs more (ROM) space but, on average, less machine instructions (basic operations) for search than the PHT.

There are additional dictionaries, e.g., the global instruction word dictionary. Dictionary words and local data variables are handled with simple hashing and small look-up tables (LUT) storing collisions by linear search. The hierarchical look-up table architecture for fast text-to-bytecode translation is shown in Fig. 8.

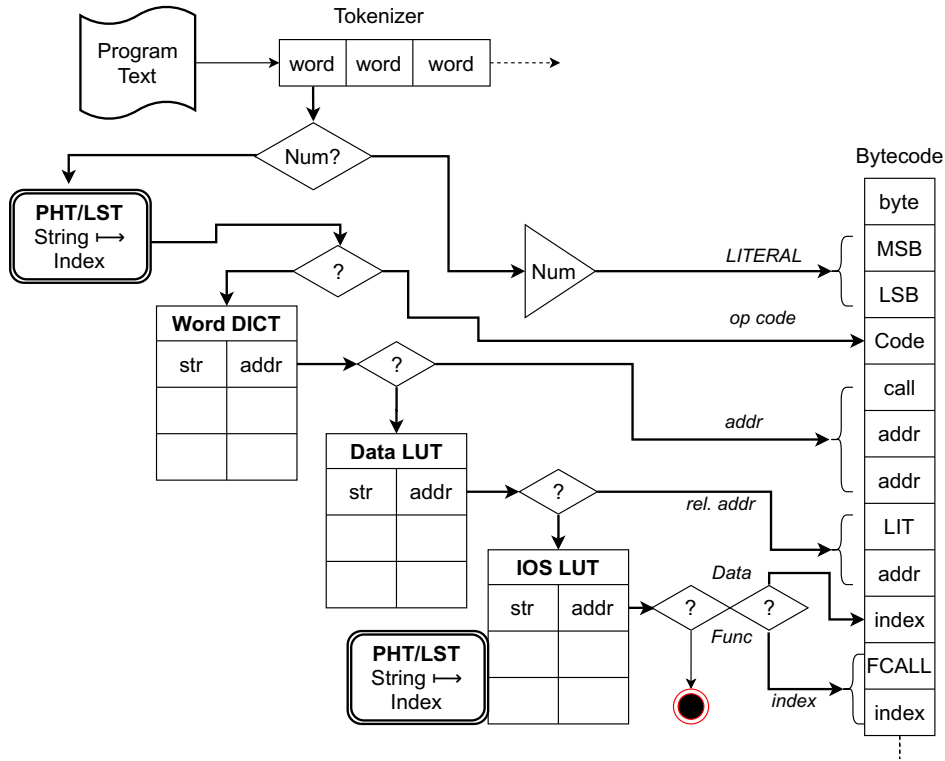

**Fig. 8.** The hierarchical compiler LUT architecture used for fast text-to-bytecode translation. After tokenization of the input text, direct op-code encoding and udpates of tables are performed.

### 3.9.1 Perfect Hash Tables

The perfect hash table is created at compile time (as part of the VM program code) from a static word list. Any modification of the word list (by adding, removing, or changing words) creates a new hash table (and word string to hash index mapping, i.e., operational code mapping). For this reason, the compiler should not be split from the VM (using a specific hash table map for the instruction decoder).

PHT has a constant run-time, but for each computation of a hash index, about $30 + n$ unity operations (arithmetic) are needed, which can be higher than LST search. A PHT look-up table required for the hash index computation consumes about $m$ bytes for $m$ words, in addition to the string check table ($m \, l_{max}$ bytes, with $l_{max}$ as the character size of the largest word from the set of words).

### 3.9.2 Linear Search Tables

LST needs at most $m$ iterations (table comparison and jump) with strings of length $m$ (independent from the number of words $n$). Each character position iteration of the word to be searched requires the iteration over a token slice, assuming about 4 average possible choices, $8m$ unity opera-tions and 8*m* look-up table accesses are required, which can be implemented with a simple FSM. The LST table consumes about 2 *n* bytes for *n* words, with an average word string length of $l_{avg}$ characters.

The LST structure is shown in Fig. 9 for an example word list. There is one subtree for each word size. The linear array consists of two-byte entries (except for the first section, which contains the start address of the first character parsing slice for each word length sub-tree). A two-byte token in the array is either a start or an inner node in the iterative search and contains the token character and the relative forward branch address to the next token slice, or it is an end (leave) node containing the word index number. A slice is iterated until a "Not Found" entry is found, terminating the search.

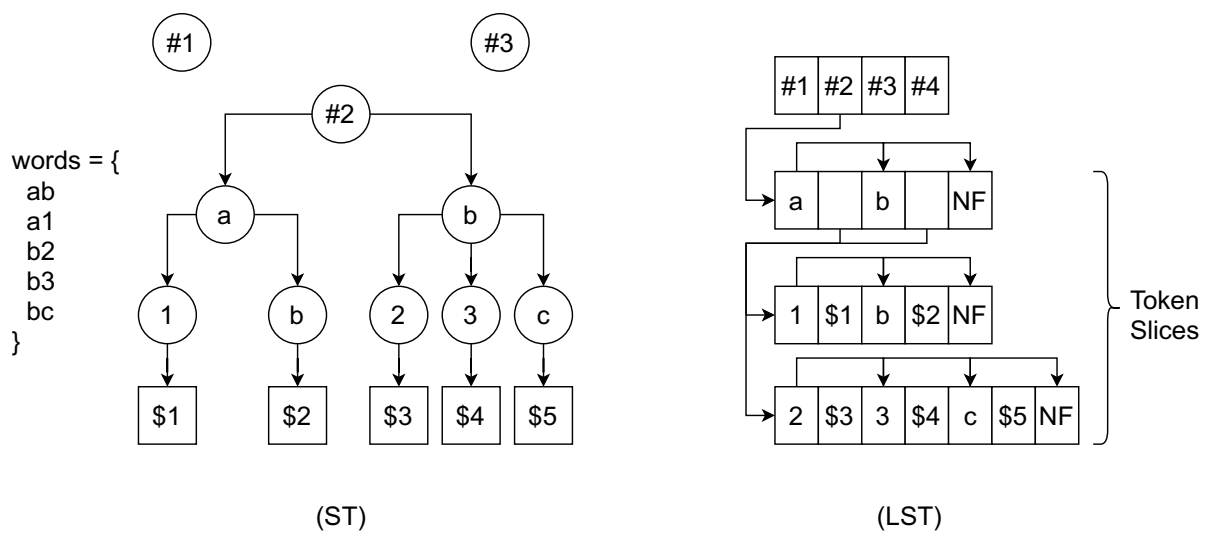

(ST)  (LST)

**Fig. 9.** (ST) A word string search tree mapping strings on unique index numbers (LST) Linear Search Table (all slices are concatenated in a linear array). There is one sub-tree for each word size (#n).

The LST size for a typical Forth instruction word table (100 words, average word length about 4 characters) is around 700 bytes, with 300 bytes accumulated characters of all words, 103 slices, minBranches = 1, maxBranches = 12, avgBranches = 2.5.

The equivalent PHT requires 128 + 700 bytes, creating a higher overhead compared with a LST. The computational logic of the search function is much simpler (4 comparators, two adders) than the hash computation in PHT (with about 10 comparators, 26 adders, 10 logical operations, and 30 register transfers, but only one look-up table access, assuming a maximal word character size of 8).

### 3.9.3 Bytecode Format

The REXA VM bytecode consists of operational code and data. Most instructions are zero-operand post-fix operations, simplifying the bytecode format significantly (and the code decoder, especially addressing hardware implementations). Def. 4 shows the REXA VM bytecode format. Literals are divided into signed short and double words (data widths of 14 and 30 bits, respectively).

**Def. 4.** REXA VM Bytecode Format (1 Byte: Post-fix operation, 2 Byte: Short word, 4 Byte: Double word, 3 Byte: Code + Address)

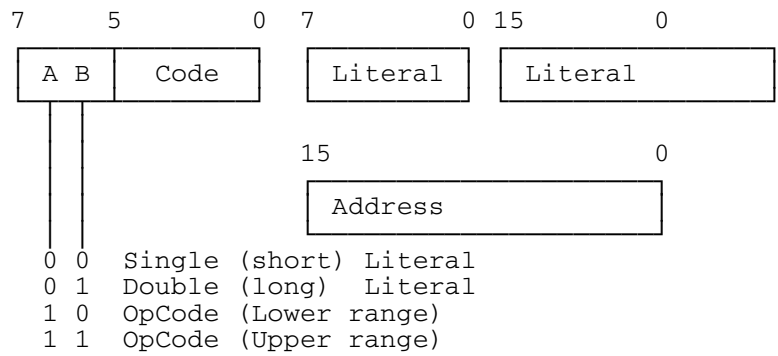

```
7       5        0 7        0 15         0
┌─┬─┬───────────┐  ┌──────────┐  ┌────────────────┐
│A│B│   Code    │  │ Literal  │  │    Literal     │
└┬┴┬┴───────────┘  └──────────┘  └────────────────┘
 │ │              15               0
 │ │              ┌──────────────────┐
 │ │              │     Address       │
 │ │              └──────────────────┘
 0 0   Single (short) Literal
 0 1   Double (long)  Literal
 1 0   OpCode (Lower range)
 1 1   OpCode (Upper range)
```

### 3.9.4 Soft- and Hardcore: Self-compiling

The compiler is either entirely implemented in the host embedded system ("hardcode") or partially by providing relevant operations like a tokenizer and core word tables, with the main part implemented in a VM program stored in a ROM or passed to the VM via communication channels. This enables compiler updates and self-compiling features and reduces native software or hardware complexity (though ROM resources are less critical in tiny embedded systems than RAM).

## 3.10  Bytecode Interpreter

The bytecode interpreter is basically the instruction decoder with a large conditional branch construct (case selector statement) mapping an operational code onto operational code statements, mostly consisting of stack manipulations. Due to the consecutive numbering of operational codes in the range { 0,1,...,opcodemax-1}, a direct branch table can be used, providing an instruction decoder with a constant run-time, which is important for real-time scheduling. Commonly, C compilers detect this feature and create a branch (look-up) table implicitly. The instruction decoder and execution unit are automatically created from the perfect hash table with a switch-case construct. The operational statements are macro definitions provided by the programmer in a separate header file. Computed *goto* statements can be created, alternatively. But not all compilers, especially commercial versions tailored for embedded systems, support computed goto statements (basically only supported by GNU compilers).

The VM bytecode interpreter is fully binary compatible among different software versions as long as the same word set is used. Any changes to the core word set (number of ops or names) invalidate binary compatibility, which is the reason for bundling the VM execution with the compiler.

### 3.10.1  Bytecode Execution Loop

The main control loop performs VM Bytecode operations in a sequential manner.The basic algorithm is shown in Alg. 1. The loop processes at least one, but maximally *steps*  instructions. The VM main loop is typically called from an application IO loop and handles various IO tasks (AD conversion, signal generation, and communication).Because the execution time for each instruction is more or less constant (within some variance), the run-time can be predicted based on a calibration, which is important to satisfy the real-time constraints of the higher IO service loop.

The next word is read from the program counter's ('pc') CS address, and the binary code is decoded, distinguishing single, double, and operational words.The main decoder consists of conditional branches, optimally implemented with a branch lookup-table based on the consecutively numbered op-code. Each operation typically consists of stack operations and the calculation of the next instruction address (program counter *pc*). Literals are operations, too. Most Forth code words are post-fix operations that consume operands from the stack store before the op is called. Some fourth code words are prefix operations that consume data from the following code words, such as *var*, *array*, and compiler-inserted (hidden) branches.

Although, Forth is a CISC high-level machine language, most instructions are simple and can be executed in constant time with comparable execution times, typically about 10-100 clock cycles. This feature enables the pre-computation of the run-time of Forth programs from the point of view of the iteration loop, satisfying time constraints. Time constraints can be strongly driven by the energy capacity limits of an autonomous sensor node.

**Alg. 1.** Bytecode execution loop template: The operational DO statement accesses stacks, CS, external functions and data, and can modify vmevent and vmerror states.

```
1:  function vmloop(pc,steps,longest) {
2:    deadline=microsec()+longest;
3:    while (step < steps && !vmerror &6 !vmevent) {
4:      if (microsec()>deadline) return pc; // Time slice exhausted
5:      instr=CS[pc];
6:      switch(DECODE(instr)) {
7:        case OP1: DO(OP1); pc=NEXT(pc); break;
8:        case OP2: DO(OP2); pc=NEXT(pc); break;
9:        ...
10:     }
11:   }
12:   return pc
13: }
```

Details of the scheduling with respect to single- and multi-tasking control flows are discussed in Sec. 6.2.

### 3.10.2 Hierarchical Scheduling under Time Constraints

In embedded systems, the VM main loop is embedded in the host application's IO loop, servicing all event- and time-based tasks. For instance, communication (reading buffers, sending data, connecting to other nodes), signal acquisition (ADC) and generation (DAC), or handling timeouts. The *vmloop* is limited by the number of VM instructions and a timeout (watchdog), ensuring the VM execution within strict time boundaries, as sketched in Fig. 10.

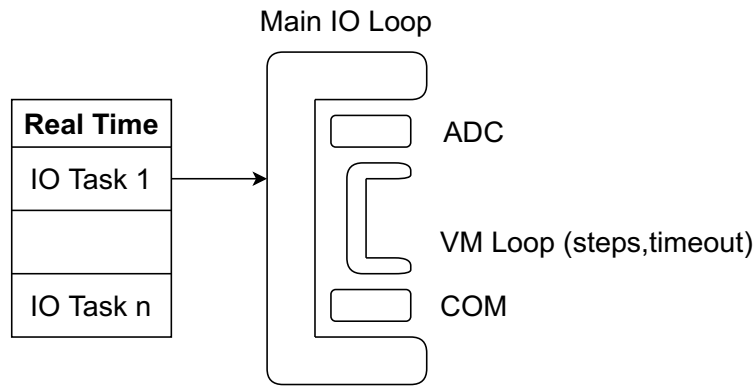

**Fig. 10.** Nested execution loops in an embedded system

## 3.11 Modular Programming

Like other interpreted programming languages, e.g., JavaScript or Lua, Forth interpreters commonly bundle the execution run-time with the compiler, especially when implementing the word dictionary that is used to resolve function call addresses during compilation. This VM system configuration provides three modes:

1. A full system mode with the Bytecode VM and the native (fast) compiler, processing code always from text blocks;

2. A stand-alone mode separating the VM from the compiler, especially suited for hardware implementations, processing only bytecode but being operationally limited;

3. A lightweight system mode with the Bytecode VM and a VM software-implemented (slower) compiler.

The program code of stack machines, e.g., Forth, is typically provided incrementally, i.e., code snippets (stored in code frames) add new function words and reference already installed function words. The separation of the compiler from the VM implies some issues requiring a small adaptation of the program code and code referencing schemes. The bundled full mode uses direct code addressing for words provided via the instruction word dictionary and indexed LUT addressing for IOS provided host application functions.

The stand-alone mode must use relative addressing, since the compiler has no knowledge of the code (address) already installed on a particular VM instance. To resolve code addresses (from the dictionary and IOS tables), there are import and export statements, shown in Def. 5. Import and export statements can be used in both modes and ensure code consistency and modularity. An import of a non-existing function will raise an exception and terminate the code-frame execution. Export statements allow the selective installation of new FORTH words in the global dictionary.

**Def. 5.** Unified Import and export statements

```
import <IDENTIFIER>
: <IDENTIFIER> .. ;
export <IDENTIFIER>
```

## 4. Tiny Machine Learning and Digital Signal Processing

Data-driven modeling is used in a wide range of classification and regression applications. Often, data-driven trained models (predictor functions) are fully trained before being used (application). Commonly, the model parameters as well as the variables are represented by floating-point data types and processed by floating-point hardware arithmetic, which is not available on low-resource microcontrollers. Basically, it is possible to use fixed-point arithmetic (e.g., with 16 bit encoded values), at least for classification tasks. Distributed machine learning is a specific class of ensemble learning based on the divide-and-conquer principle. Each node provides a local state estimation or classification based solely on local data, which is then globally fusioned to a global state.Assuming such an architecture, predictive classification (and possibly regression) is appealing for implementation on the sensor node level and directly processed by the microcontroller or FPGA processing unit, a concept known as "tiny machine learning."

One prominent sub-class of predictive data-driven models are Artificial Neural Networks (ANN), which are basically non-linear function graphs.

An ANN can be organized in layers, and each layer consists of a given number of functional nodes (neurons). Each functional node  performs a data fusion by summing the products of all input variables $\boldsymbol{x}$ (vector) with a weight parameter vector $\boldsymbol{w}$. Finally, the resulting scalar value $t$ is passed to a commonly non-linear transfer function $g(t)$, which provides the node's output.For the computation of one node, vector operations are required.

To compute (apply) an ANN, only some specific vector arithmetic operations and a unified vector and matrix data structure are required. A challenge is the reduction in value of resolution and precision. ANNs are typically trained using floating-point arithmetic (at least with a single 32-bit precision).The VM addressed in this work supports only 16- and 32-bit integer arithmetic. The transformation of already trained networks into integer interval arithmetic requires additional scaling vectors and scaling operations.

The ANN can be functionally decomposed into the following vector and matrix operations:

$$f : R^n \rightarrow R^p \approx I^n \rightarrow I^p$$
$$f = g \circ f_{l-1} \circ f_{l-2} \circ .. \circ f_1$$
$$f_i(\vec{x}) = a\left(\hat{w}_i \vec{x} + \vec{b}_i\right)$$
$$g(\vec{z}) = \begin{cases} z & \text{regression} \\ \frac{1}{1+e^{-z}} & \text{binary classification} \\ \frac{e^{z_j}}{\sum_k \left(e^{-z_k}\right)} & \text{multi-classification} \end{cases} \tag{2}$$

All functions $f_i$ (representing one layer) and the output function $g$ use matrix and vector operations, which can be implemented in software as well as hardware and computed directly with integer arithmetic. Only the activation (transfer) functions (e.g., sigmoid or soft-max) require approximated fixed-point (integer) implementations of the real-valued functions, typically using a combination of piecewise multi-point regression and look-up tables. Not fully connected ANNs are computed in the same way as fully connected ANNs, but they produce sparse vectors and matrices, resulting in a large number of null (useless) operations.

The vector operations in REXA VM's hardware architecture can be parallelized, balancing resource occupation and speed. The ARM Cortex M0-M3 processors do not provide parallel vector operations (such DSP operations were added first in generation 4).

## 4.1 Signal Interface

A sensor node processes sensor data, which is commonly sampled by the node itself using analog-digital conversion (ADC). Active measuring techniques necessitate the generation of a stimulus, which is typically controlled by the sensor node (e.g., a digital-to-analog converter (DAC)).VM programs can access the signal acquisition layer by using a signal device interface provided by the sensor node host application via the IOS. Sampled sensor data is stored in a dedicated buffer, commonly filled automatically during the sampling phase via direct memory access (DMA). The sample buffer can be directly accessed by the VM or at program level. DMA sampling with triggering, e.g., on a specific threshold level, typically utilizes a ring buffer memory architecture. Reading the sample buffer must also be done cyclically in this case, beginning at a specific top buffeer position.Due to hard resource constraints, the sample buffer is also used for digital signal processing, e.g., applying filters to the data in-place. The sensor and actuator API is summarized in Tab. 3.

| Operation | Operands | Description |
|---|---|---|
| adc | ( trigmode depth amp-Gain sampleFreq device -- ) | Starts an AD conversion with a specific trigger mode, sample buffer depth (number of samples in kS), pre-amplifier gain if selectable, sample frequency (in kSPS), and the device or channel selector. The conversion runs concurrently to the VM processing, the op does not suspend program flow. |
| dac | ( wave interval ampl sampleFreq device -- ) | Starts a DA conversion with data from a specific wave table, continous periodically (like sine waves) or single-shot with time delay, output amplitude, sample frequency, and the device or channel selector. The generation runs concurrently to the VM processing, the op does not suspend program flow. |
| sampled | ( -- addr ) | Pushes the address (IOS index) of the AD conversion status variable on the stack. |
| samples | ( -- addr ) | Pushes the address (IOS index) of the sample buffer on the stack. |
| sample0 | ( -- addr ) | Pushes the address (IOS index) of the sample buffer top offset variable on the stack. |
| wave | ( -- addr ) | Pushes the address (IOS index) of the programmable wave buffer on the stack. |

**Tab. 3.** Signal acquisition and generation API

The following example Ex. 1 demonstrates an AD conversion with synchronous waiting (line 3) for completion of the conversion (or timeout) and final post processing finding the peak value and position (lines 8-12).

**Ex. 1.** Synchronous AD conversion with post processing

```
1: const FREE 10 const SINGLE 4 const HIGH 1
2: FREE 1 HIGH 100 0 adc ( Start ADC )
3: 1000 1 sampled await ( Suspend task )
```

```
 4:  <0 if error endif
 5:  var peak 0 peak !
 6:  var offset sample0 read !
 7:  var pos
 8:  1024 0 do  ( Iterate over sample buffer )
 9:    offset @ samples read
10:    dup peak @ > if peak ! i pos ! else drop endif
11:    offset @ 1 + 1024 mod offset !
12:  loop
13:  ." Peak: " peak @ ." at " pos @ . cr
```

## 4.2  Digital Signal Processing

The set of DSP operations is provided via the FIOS layer API and can be extended by the host application. Only fixed-point integer arithemtic is supported. The input and output scaling of arithmetic and numerical functions is fixed. Basic operations required for typical signal processing and analysis tasks are provided, as shown in Tab. 4.

| Operation | Operands | Description |
|---|---|---|
| sin | ( x -- y ) | Integer discrete sine function (x/y scale: 1000) |
| log | ( x -- y ) | Integer discrete logarithmic(base 10) function (scale: x 10, y 1000) |
| sigmoid | ( x -- y ) | Integer discrete sigmoid function (x/y scale: 1000) |
| relu | ( x -- y ) | Integer discrete rectifying linear unut function (x/y scale: 1000) |
| hull | ( vecaddr vecoff veclen k -- ) | In-place signal hull computation using a low-pass frequency filter |
| lowp | ( vecaddr vecoff veclen k -- ) | In-place signal low-pass frequency filter |
| highp | ( vecaddr vecoff veclen k -- ) | In-place signal high-pass frequency filter |

**Tab. 4.** Basic Digital Signal Processing functions provided by the external DSP library via FIOS

Trigonometric functions and functions composed of trigonometric functions are implemented with segmented linear and non-linear look-up tables. For example, the error of the discrete sigmoid function is always less than 1%, while only requiring 30 bytes of LUT space and less than 10 unit operations, as shown in Alg. 2. These software functions can be immediately implemented in hardware, too. The LUTs are computed with Alg. 3.

**Alg. 2.** Range-segmented and LUT-based implementation of the sigmoid function with less than 1% approximation error (using approximated LUT-based log10 function)

```
 1:  static ub1 sglut13[] = { <24 values> };
 2:  static ub1 sglut310[] = { <6 elements> };
 3:  // y scale 1:1000 [0,1], x scale 1:1000
 4:  sb2 fpsigmoid(sb2 x) {
```

```
 5:    sb2 y;
 6:    ub1 mirror=x<0?1:0;
 7:    if (mirror) x=-x;
 8:    if (x>=10000) return mirror?0:1000;
 9:    if (x<=1000) {
10:       y = 500+(((x*231)/1000));
11:       return mirror?1000-y:y;
12:    } else if (x<3000) {
13:       ub2 i10 = ((fplog10((x/5)|0)/2))-65;
14:       y = ((sb2)sglut13[i10])+731;
15:       return mirror?1000-y:y;
16:    } else {
17:       ub2 i10 = ((fplog10((x/10)|0)/10))-14;
18:       y = ((sb2)sglut310[i10])+952;
19:       return mirror?1000-y:y;
20:    }
21:    return 0;
22: }
23: static ub1 log10lut[] = { <100 values> }
24: // x-scale is 1:10 and log10-scale is 1:100
25: sb2 fplog10(sb2 x) {
26:    sb2 shift=0;
27:    while (x>=100) { shift++; x/=10; };
28:    return shift*100+(sb2)log10lut[x-10];
29: }
```

The LUT tables can be computed as follows:

$$\text{log10lut} = \left\{ int\left(\log_{10}\left(\frac{i}{10}\right) \cdot 100\right) : i \in \mathbb{I}, 0 \le i \le 99 \right\} \tag{3}$$

The *fpsigmoid* function LUTs are computed iteratively using the *fplog10* function, described by the following pseudo code algorithm Alg. 3 (accuracy is plotted in Fig. 11:

**Alg. 3.** Computation of the LUTs for the fixed-point sigmoid function

```
 1: sglut13 := []
 2: for x=1 to 2.95 step 0.05 do
 3:    i10 := int(fplog10(int(x*1000/5))/2)-65
 4:    if sglut13[i10] = undefined then
 5:       sglut13[i10] := int(sigmoid(x)*1000)-731
 6:    endif
 7: done
 8: sglut310 := []
 9: for x=3 to 9.9 step 0.1 do
10:    i10 := int(fplog10(int(x*1000/10))/10)-14
11:    if sglut310[i10] = undefined then
12:       sglut310[i10] := int(sigmoid(x)*1000)-952
13:    endif
14: done
```

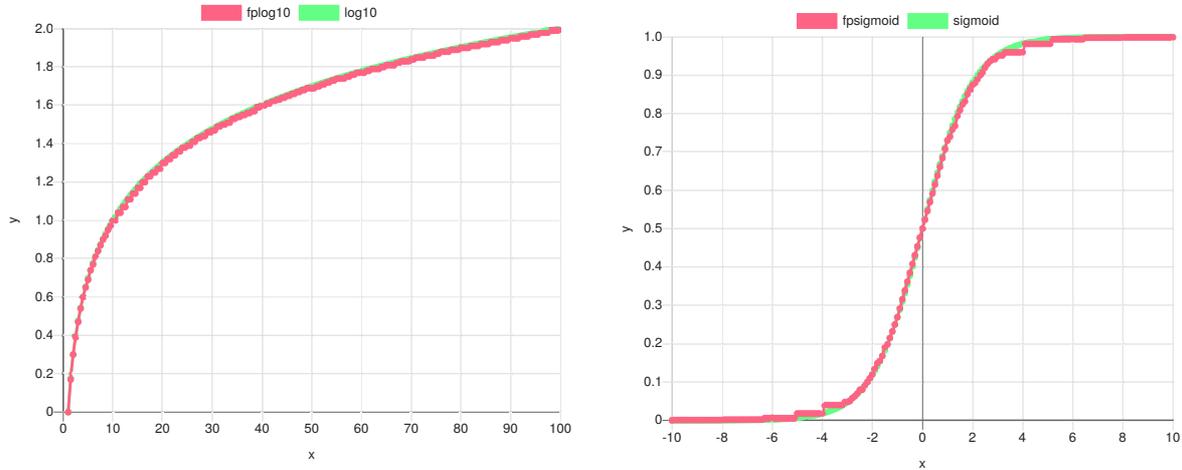

**Fig. 11.** Accuracy of fixed-point approximations for *log10* and *sigmoid* functions

### 4.3 Artificial Neural Network (ANN)

An ANN consists of two parts:

1. The data, i.e., for parameter, input, and output variables;

2. The structure and functions processing the data.

For the sake of simplicity, fully connected networks are assumed, but any irregular network structure is a sub-set of a fully connected structure and can be used with the following operational architectures, too. In contrast to common ANN software frameworks, REXA VM provides only core vector operations. The parameter data is embedded in a code frame by using the initialized *array* constructor. Both parameter and input/output data can be stored in the program code frame, shown in the next section.

#### 4.3.1 Vector Operations

All the basic operations you need to implement ANNs and perform forward activation computations are:

1. Element-wise vector operations (*vecmul*: array → array);

2. Dot-product operation performing a sum of product data fusion (*dotprod*: array × array → number );

3. A folding operation for node layer computations (*fold*: array × array → array )

4. And for fixed-point integer arithmetic scaling operations are required (*scale*: array × array → array ).

The core set of vector operations provided by the REXA VM supporting integer arithmetic ANN computations is summarized in Tab. 5.

Vector operations always operate on single data words (16 bit), but internally 32 bit arithmetic is used to avoid overflows. To scale to signed 16 bit integer, some of the operations use a scale factor or scale factor vector (negative scale values reduce, positive expand the values by the scale factor) to avoid overflows or underflows in following computations, similar to scaled tensors in [16,17].

| Operation | Operands | Description |
|---|---|---|
| `array`<br>`<ident>`<br>`<#cells>` | - | Allocates a data array in the code segment |
| `vecload` | `( srcvec srcoff dstvec -- )` | Loads a data array into another array buffer. The source can be any external data provided by the IOS or internal embedded data. The source array (plus optional offset) must have at least the size of the destination array |
| `vecscale` | `( srcvec dstvec scalevec -- )` | Scales the source data array with scaling factors from the scale array and stores the result in the destination array. Negative scalign values reduce, positive values expand the source data values. |
| `vecadd` | `( op1vec op2vec dstvec scalevec -- )` | Adds two vectors element-wise with an optional result scaling (value 0 disables scaling). Both input and the destination vectors must have the same size. |
| `vecmul` | `( op1vec op2vec dstvec scalevec -- )` | Multiplies two vectors element-wise with an optional result scaling (value 0 disables scaling).Both input and the destination vectors must have the same size. |
| `vecfold` | `( invec wgtvec outvec scalevec -- )` | Performs a folding operation *ivec* × *wgtvec*. The weights vector *wgtvec* must have the size $\|invec\| * \|outvec\|$. |
| `vecmap` | `( srcvec dstvec func scalevec -- )` | Maps all elements from the source array onto the destination array using an external (IOS) or internal (user-defined word) function, e.g., the sigmoid function. |

**Tab. 5.** Basic vector ANN functions operating on embedded or external array data (e.g., the sample buffer)

The operations are defined by the following formulas:

$$vecmul\left(\vec{a}, \vec{b}\right) = (a_1 \cdot b_1, a_2 \cdot b_2, .., a_n \cdot b_n)^T$$

$$dotprod\left(\vec{a}, \vec{b}\right) = \sum_{i=1}^{n} a_i \cdot b_i$$

$$fold(\vec{a}, \hat{c}) = \left(\sum_{i=1}^{n} a_i \cdot c_{i,1}, \sum_{i=1}^{n} a_i \cdot c_{i,2}, .., \sum_{i=1}^{n} a_i \cdot c_{i,n}\right)^T \quad (4)$$

$$map(\vec{a}, f) = (f(a_1), f(a_2), .., f(a_n))^T$$

$$n = |\vec{a}| = \left|\vec{b}\right|$$

$$|c| = n \cdot m$$

### 4.3.2 ANN Example

The following example code shows how simple it is to implement ANNs in REXA Forth, as shown in Ex. 2, by implementing a [4, 3, 2] network layer node configuration.The network parameters (lines 3-26) are embedded in the code and are used by the network forward activation function *forward* (lines 32-48). The computations are performed by using the pre-defined universal vector operations, introduced in the previous section. The vector operations determine the size parameters of the vectors (or matrix) automatically. Impressive performance results are presented in Sec. 8.

**Ex. 2.** ANN consists of 4 input variables and neurons, one hidden layer of 3 neurons, and two output neurons. The ANN is implemented entirely in one code frame (about 400 bytes).

```
 1: ( Signed 16 bit integer type arrays )
 2: ( Input Layer )
 3: array input  4
 4: array biasI { -2 15 0 1 }
 5: array wghtI { 10 -15 10 2 }
 6: array scaleI { 10 -10 2 5 }
 7: array activI 4
 8:
 9: ( Hidden Layer )
10: array wghtH1 {
11:   10 -5 4 2 ( Neuron 1 )
12:   0 1 1 0    ( Neuron 2 )
13:   5 -2 -2 0 ( Neuron 3 )
14: }
15: array biasH1 { -4 5 10 }
16: array scaleH1 { -2 10 -8 }
17: array activH1 3
18: array wghtO {
19:   2 5 9
20:   6 1 0
21: }
22:
23: ( Output Layer )
24: array biasO { -1 1 }
25: array scaleO { -2 10 }
26: array output 2
27:
28: ( Load input vector from sample buffer starting from offset [100] )
29: sample 100 input vecload
30:
31: ( Forward activiation of network )
32: : forward
33:   ( Evaluate input layer --   )
34:   input wghtI actI scaleI vecmul
35:   ( Add bias )
36:   actI biasI actI 0 vecadd
37:   ( Apply activation function w/o scaling )
38:   actI acI $ sigmoid 0 vecmap
39:   ( Compute hidden layer activations )
40:   actI wghtH1 activH1 scaleH1 vecfold
41:   activH1 biasH1 activH1 0 vecadd
42:   ( Apply activation function w/o scaling )
43:   activH1 activH1 $ sigmoid 0 vecmap
44:   ( Compute output layer activations )
45:   activH1 wghtO output scaleO vecfold
46:   output bias output 0 vecadd
47:   output output $ sigmoid 0 vecmap
48: ;
49: forward
50: output vecprint cr
```

*51: ( Done )*

---

The initialized vectors are stored in-place in the code frame, the non-initialized vectors are stored at the end of the code frame (extending the code frame by the compiler).

## 4.4 Decision Trees

Decision trees, as lightweight predictor models well suited for tiny embedded systems, can be efficiently stored in Linear Search Tables (LST), as introduced earlier for compiler parsing. Decision trees consist of nodes associated with input variables $x_j$ or output variables $y_k$ (and specific outcomes of a prediction). Directed edges connecting nodes are functional evaluations of a node variable. There are three basic operations: Binary relation (</>), equality (=), and nearest value approximation (≈). The data format is shown in Def. 6. Each slide starts with the input variable to be evaluated (or target for output), the operation applied to choices, a field specifying the number of choices, and value-branch pairs.

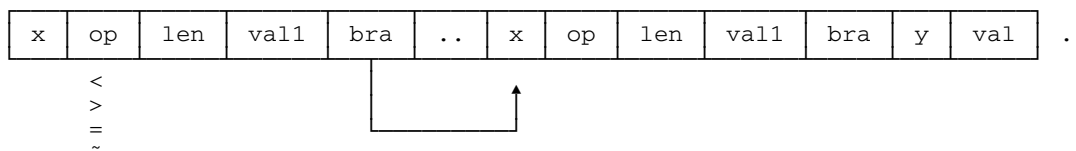

**Def. 6.** Format of a Linear Serach Tree (LST) implementing a decision tree

## 5. Software and Hardware Implementations

The REXA VM can be implemented in three ways:

1. Software implementation as a library, including VM and compiler (or VM only), and integrated in a host application program (sensor node software) using shared-memory communication;

2. Co-processor: Hardware implementation of the VM with a multi-FSM RTL architecture targeting FPGAs (or ASICs) connected to the host software via a communication bus or serial link (co-processor);

3. Stand-alone: Hardware implementation of the VM with a multi-FSM RTL architecture targeting FPGAs (or ASICs) with communication and signal processing modules.

Both implementations are fully binary compatible, as long as the same word set is used. Any changes to the core word set (number of ops or names) invalidate binary compatibility.

## 5.1 Code Generators

The VM and compiler should be implemented on a wide range of devices, processor architectures, both in software and in hardware (FPGA-RTL). The instruction set architecture is not static and is customizable. For the sake of flexibility and optimal matching of host processing architectures, the source code of the VM and the compiler are split into two classes:

1. Static source code modules targeting a specific programming language (including HDL), e.g., the implementation of the stacks, the dictionary, and shared compiler parts;

2. Source code is generated dynamically by source code generators targeting a specific programming language (including HDL), and based on configuration files, such as a core word list file.

## 5.2 Software Implementation

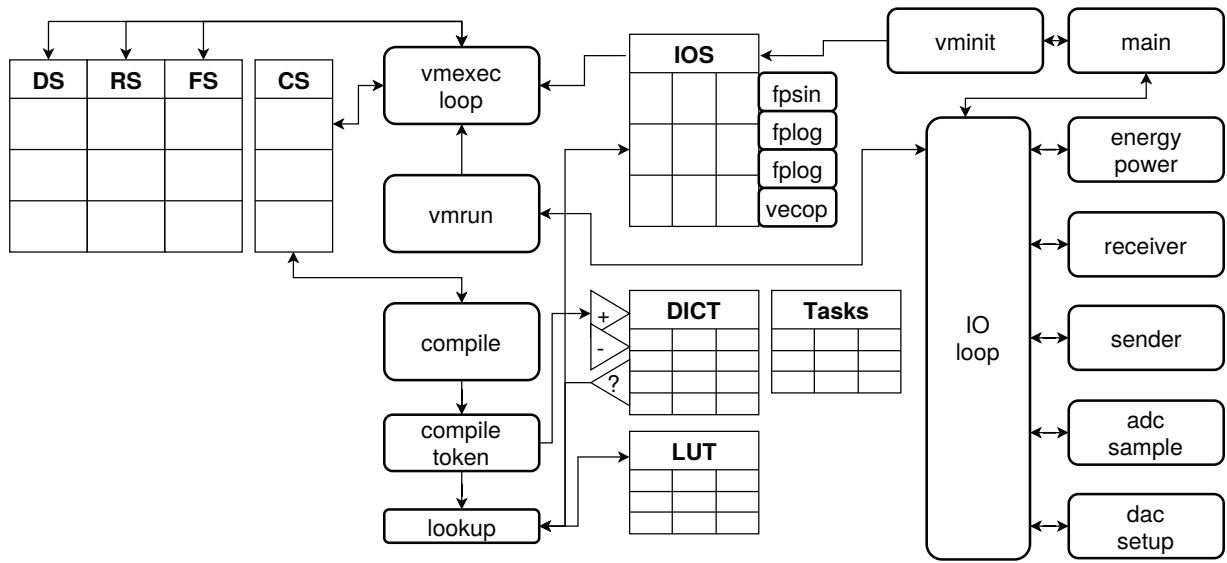

**Fig. 12.** Functional composition of the software architecture of REXA VM (simplified)

## 5.3 Hardware Implementation

The hardware implementation of REXA VM maps functional units of the VM onto sequential state-based RTL processes with a control path (Finite State Machines) and a data path, as shown in Fig. 13. The CS and the stacks are implemented with block RAM components. The hardware implementation can exploit parallel processing by replicating the central VM execution loop and the stack memories (two threads shown in Fig. 13). Each VM thread instance is controlled by its own task and resource scheduler. One central scheduler manages all VM instances and distributes code frame execution to VM thread instances.

VM thread instances share the CS and the IOS scheduler, gaining access to device drivers and co-processors, e.g., the DSP processor implementing vector operations used for ML computations.

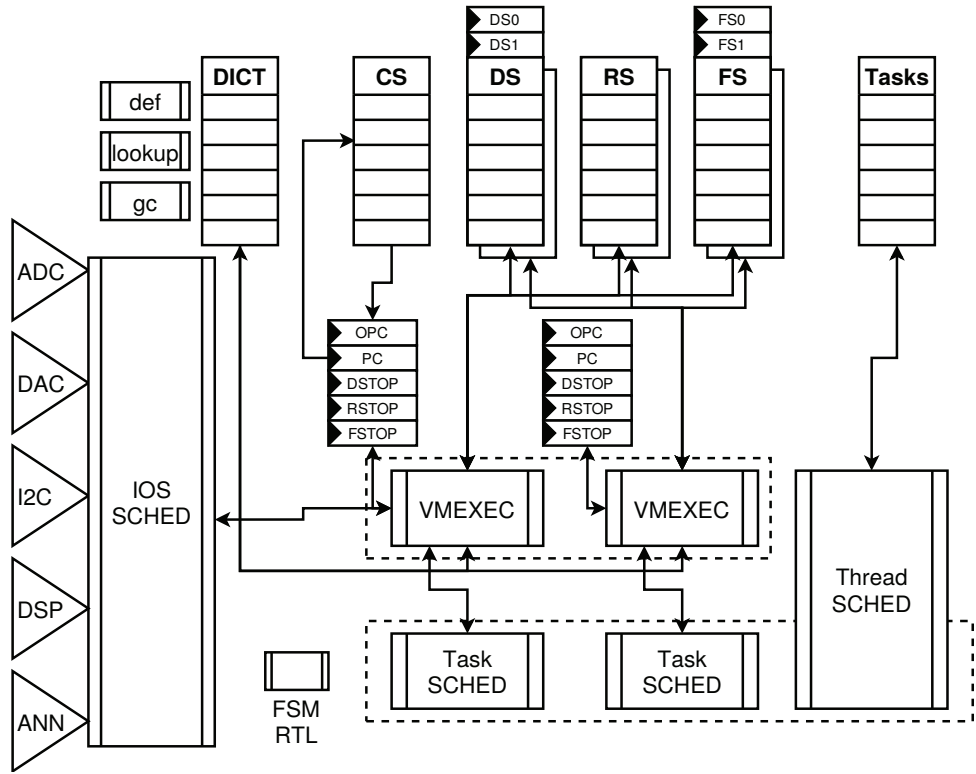

**Fig. 13.** Multi-process hardware architecture of REXA VM (simplified)

The hardware architecture utilizes control path parallelism by VM execution loop replication, hardware process pipelining using queues, and data path parallelism by executing arithmetic operations in parallel and by using fixed-width parallel basic vector operations for the DSP extension.

The VM execution loop is not directly activated. Instead, the VM loop is connected via a control event queue, with the upper level scheduler providing synchronisation. The control queue exchanges task tokens. A task token executes a specific task for a given maximum number of steps and time. If the task terminates, is suspended, or a timeout occurs, a token is sent back to the scheduler via a status event queue.

## 6. Real-time Scheduling under Energy Constraints

A sensor node, in particular, has to process a set of tasks characterised by different priorities, arrival times, deadline, and execution times:

1. Signal sampling and generation (triggering)

2. Event detection

3. Communication (on-chip, on-board, or externally wireless)

4. Computation

5. Energy Management (energy savings and optimisation)

6. Service requests and processing (deadlines)

7. Watchdog (sensor and node failure detector)

These tasks must be scheduled under time and performance constraints. Assuming only one physical control path (one processor), the tasks must be scheduled in slices by one main scheduling loop. Self-powered sensor nodes introduce additional energy constraints, which are discussed in the next section. The tasks can be classified as event-based (short-running), data-driven (long running), and communication (event- and data-driven) tasks.

## 6.1 Energie-driven Scheduling

A self-powered sensor node obtains energy from its surroundings (the source), and an energy harvester provides power $PS(t)$. This power, which is given by P, can be consumed by the embedded system electronics and sensors (sink). If the node lacks an energy storage device (capacitor), the run-time of the embedded system is limited by satisfying the condition $P_S/P_D > 1$ and the time interval $t_S = [t_1, t_2]$ of the presence of the energy field, as shown in Fig. 14. If no energy storage is available and no pre-charging is possible, the run-time must be predicted by the expected power cycle. In this work, NFC is used to supply the sensor node. The average power cycle is about 10-30 seconds. The power loss can result in a premature interruption of the current task processing, as shown in Fig. 14 (a), not occur in the case where there is an energy storage with known energy capacity (the node can predict the run-time by itself), as shown in Fig. 14 (b). Power supply and energy capacity are important parameters for real-time task scheduling, with the goal of executing and completing at least high-priority tasks.

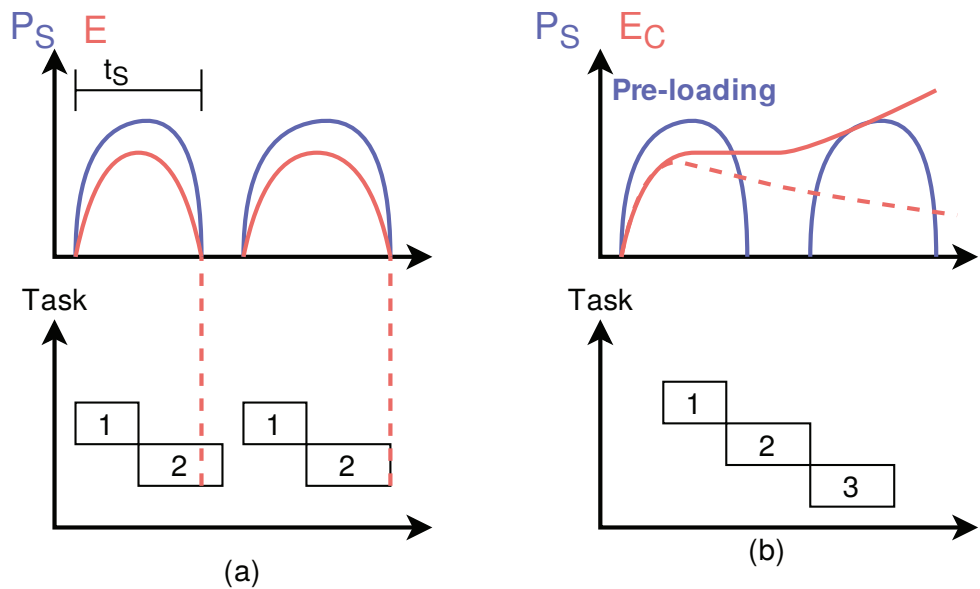

**Fig. 14.** Task scheduling under real-time and energy constraints (a) Node without an energy storage (b) With energy storag

As long as there is no clock frequency scaling, the consumed power $P_D(t)$ is nearly constant, and the consumed energy depends only on the computation time.

According to [9], a Lazy Scheduling Algorithm (LSA) is used for task scheduling under energy constraints. Assume there is a set of Tasks $T = \{ p_1, p_2, .., p_n \}$ to be managed at some time by the processing system (here the VM and VM-external IO tasks not visible to the VM). A sub-set of the tasks, $T_r \subseteq K$, contains tasks ready to run and to be scheduled. Another sub-set of the tasks $T_{ev} \subseteq T$, contains tasks waiting for events (not ready).

A source of energy absorbs ambient energy at time $t$ and transforms it into electrical power $P_S(t)$. This power can be stored in a storage device with energy capacity $C$. The stored energy is $E_C<C$. On the consumer side, a computing device employs power $0 \leq P_D(t) \leq P_{max}$ that is drained from the storage to perform jobs with deadlines $d_i$, energy demands $e_i$, and arrival times $a_i$. Pre-emption is permitted, and it is assumed that there is only one task running at a time. The effective starting time $s_i$ and finishing time $f_i$ of a task $i$ are dependent on the scheduling strategy used: A task starting at time $s_i$ will finish as soon as the required amount of energy $e_i$ has been consumed by the task. LSA degenerates to an "earliest deadline first" (EDF) policy, if $C=0$.

The basic scheduling algorithm is shown in Alg. 4. If the current time $t$ equals the deadline $d_j$ of some arrived but not yet finished task $k_j$, then finish its execution by draining energy ($e_j - E_j (a_j, t)$) from the energy storage $C$ instantaneously (lines 12-18). In addition to arrival and deadline times, each task has a priority $r_j$. Selecting a task $j$ bases on:

1. Priority;

2. Arrival and deadline times;

3. Energy deposit or exepcted power cycle end (by stochastic prediction and sensing);

4. Expected run-time of a task (by estimation and profiling) $t_j$;

5. Progress of task execution (e.g., a task can be pre-empted or suspended waiting for an event).

It is assumed that the power consumption of the processor cannot be adjusted (e.g., by changing the core voltage and/or the clock frequency), which is set to a pre-determined value of $P_{d1}$. The energy deposit is monitored (if there is any energy storage) and updated by VM processing (or external IO task processing). Either the number of processed VM instructions (*steps*) or the processing can be used. High resolution clocks with deadlines are not always available in tiny embedded systems; therefore, an estimation is used. If there is no storage, then $E$ is set to zero. Instead, if there is a $P_S > 0$, this measure is used. There is predictive system run-time (under the assupmtion of $P_D(t)=P_{d1}=$const) $t_s$.

---

**Alg. 4.** Modified LAS scheduling algorithm for energy constrained embedded systems

```
 1: E   := E₀
 2: P_d1 := const mW Power
 3: t₁   := µs
 4: function schedule (T_r)
 5:    compute minimal t_s
 6:    Q <= sort{ min{d_i} ∧ max{r_i} : i ∈ T_rdy }
 7:    do
 8:       j <= removehead (Q)
 9:    while t_j > t_s
10:    d    <= d_j
11:    task <= p_j
12:    if task is IO task then
13:       process task with power P_D(t)=P_d1
14:       E := E - P_d*Δt
15:    else
16:       compute current e_max and estimated running time in steps
17:       process task with power P_D(t)=P_d1 via vmloop(steps,e_max)
```

```
18:      E := E - P_d*steps*t_1
19:   end
20:   t <= current time
21:   ∀ k ∈ T_ev: if t ≥ a_k then add p_k to T_r
22:   if t ≥ f_j then remove p_j from T_r
23:   if t ≥ d_j then
24:     if task is IO taks then
25:       process task with power P_D(t)=const
26:     else
27:       e <= (e_j - E_j(a_j ,t))
28:       process task with power P_D(t)=const via vmloop(steps,e)
29:     end
30:     remove p_j from T_r
31: end
```

## 6.2  Predictive Task Scheduling

It is assumed that node tasks execute functions with a specific run-time behaviour: One-time, scheduled (pre-emptive), or periodically. The execution of VM programs and tasks is a subset handled by the main IO scheduling loop. Static and dynamic priorities are assigned to tasks (and classes: event, computation, communication). The entire run-time of a task or a specific instruction word is not known in advance. Word and task profiling by the VM enables run-time prediction (number of VM instructions required for the entire word, average until the scheduling point). Finally, for systems with short power-on cycles, check-pointing can be used to save the VM state.

In a tiny embedded system, there is one main IO scheduling loop (endless looping) that performs the allocation of limited computing resources, especially addressing on/sleep/off phases due to a temporal lack of energy. Energy management and monitoring can be part of a real-time scheduler. The application-specific main loop can keep track of high-priority IO tasks without compromising VM execution or using a real-time scheduler.

## 6.3  Minimal Single-Tasking

The scheduling and calling of the inner VM instruction loop is rather simple if there is only one control context (task), as shown in Alg. 5.

**Alg. 5.** Minimal single-tasking VM execution flow

```
1: if (vmevent.timeout && now >= vmevent.timeout) {
2:   pc=-pc;
3:   vmevent.timeout=0;
4: }
5: if (vmevent.v && *vmevent.v == vmevent.c) {
6:   pc=-pc;
7:   vmevent.v=NULL;
8: }
9: if (pc>=0) pc=vmloop(code,pc,steps);
10: if (pc<0) {
11:   /* suspended */
12: }
```

## 6.4 Multi-Tasking

Assuming there is only one processor (core), multi-tasking is simulated with a scheduler. True multi-threading with concurrent and parallel control flows requires a set of separate stacks, one for each thread. But even scheduled multi-tasking (one control flow at a time) requires this. A scheduler switches (multiplexes) the VM loop between different tasks and must swap program and stack pointers each time a thread schedule occurs. Task scheduling can be either preemptive (as supported by REXA VM) or cooperative (via blocking opertations).

Event-based scheduling can be done efficiently and has low overhead. One major drawback of multi-threading is the partitioning of all stacks, providing one partition for one thread, compromising the resource efficiency of stack VMs, and increasing the memory requirement significantly because all memory is allocated in advance.

Because different tasks are associated with different contexts and environments (stacks, program pointers, events), a more complex scheduler is required. The multi-tasking version of the scheduling operation is shown in Alg. 6. This is a highly optimized version that reduces the number of operations on task scheduling and event handling significantly by introducing a tasks mask (for ready, time-out, or event awaiting tasks).

**Alg. 6.** Multi-tasking VM execution flow (without energy-driven real-time scheduling)

```
 1: #define TASKREADY    0b11111111111111111111111111111111
 2: #define TASKTIMEOUT 0b01010101010101010101010101010101011
 3: #define TASKEVENT   0b10101010101010101010101010101010101
 4: sb2 DS[STACKSIZE*MAXTASKS]; // ...
 5: schedule: if (vmtaskmask) {
 6:   // some thread(s) to manage
 7:   ub4 mask=0x3;
 8:   sb1 next=-1;
 9:   for(i=0;i<vmtasktop;i++) {
10:     if ((vmtaskmask&mask)==TASKREADY && next==-1) {
11:       // Schedule: Lowest priority
12:       vmtaskmask=(vmtaskmask & ~mask);
13:       next=i;
14:     } else if ((vmtaskmask&mask)==TASKTIMEOUT && vmtaskss[i].timeout<now) {
15:       // Time Event: high priority
16:       vmtaskmask=(vmtaskask & ~mask);
17:       vmtasks[i].timeout=0;
18:       next=i;
19:       break;
20:     } else if ((vmtaskmask&mask)==TASKEVENT   && *vmtasks[i].v==vmtasks[i].c) {
21:       // IO Event: highest priority
22:       vmtaskmask=(vmtaskmask & ~mask);
23:       vmtasks[i].v=NULL;
24:       next=i;
25:       break;
26:     }
27:     mask=mask<<2;
28:   }
29:   if (next>=0) {
30:     pc=vmtasks[next].pc; dstop=vmtasks[next].ds; rstop=vmtasks[next].rs;
31:   };
```

```
32: }
33: if (pc>=0) pc=vmloop(code,pc,steps);
34: if (pc<0) {
35:   /* suspended, update threadmask */
36:   ub4 mask;
37:   if (vmtasks[vmtaskcur].timeout) mask= 0b01 << vmtaskcur;
38:   else if (vmtaskss[vmtaskcur].v) mask= 0b10 << vmtaskcur;
39:   else mask= 0b11 << vmtaskcur;
40:   vmtaskmask|=mask;
41:   vmtasks[i].pc=-pc;vmtasks[i].ds=dstop;vmtasks[i].rs=rstop;
42: }
```

## 7. Use-cases

### 7.1 Software Sensor Node

The original motivation for virtualization and the deployment of VMs on tiny microcontrollers was the wireless material-integrated sensor node, which is embedded between two layers of a fiber-metal laminate plate and developed in the DFG research group 3022 for automated diagnostics of hidden damages in fiber-metal laminates. The sensor node is supplied with power via NFC only. The energy harvester is able to deliver up to 15 mW of continuous power, constraining the selection and operation of the electronics parts significantly, as shown in Fig. 15. After the sensor node is integrated into the plate, no software updates or maintenance can be done. The communication with the microcontroller takes place only via the NFC tag. The communication is bidirectional, originally via the NFC tag's EEPROM (code and data). Alternatively, the wireless communication can be realized directly by using a write-through mode (the lifetime of the EEPROM is limited to about 1000-10000 write cycles) and directly writing message data to the microcontroller. The sensor node uses an ARM Cortex STM 32 L031 device, an NFC tag with a power supply, and a pre-amplifier for piezoelectric sensors. The features are summarized in Tab. 6.

| Feature | Value |
|---|---|
| Size | 17 × 17 mm (without antenna and sensor) |
| Sensors | Piezoelectric transducer, MEMS sensor, temperature, radio field strength |
| Components | ADC (1 MSPS, 8-12 bit resolution), pre-amplifier, power regulator, NFC tag, ARM Cortex M0 STM32L031 microcontroller with 8 kB RAM and 32 kB ROM |
| Communication | Wireless, NFC (13.56 MHz), up to 100 kb/s |
| Energy | Energy harvesting by NFC tag, up to 15 mW continous power |
| Software | REXA VM (CS=1024, DS=512, RS/FS=32, only single-precision data) |

**Tab. 6.** Features of the material-integarted wireless sensor node

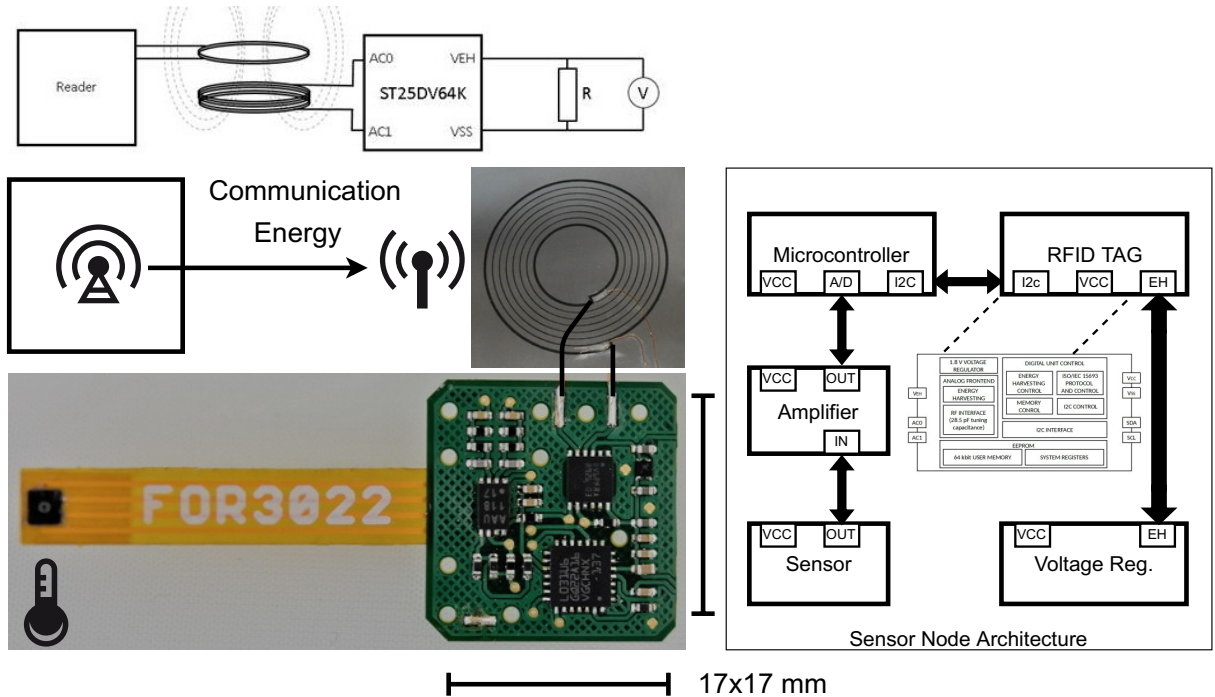

**Fig. 15.** ARM Cortex M0 sensor node (STM32L031) for material-integrated GUW sensing with NFC for energy transfer and bidirectional communication implementing the REXA VM (8 kB RAM, 32 kB ROM)

## 7.2 Hardware Sensor Node

The hardware sensor node is an alternative implementation consisting of ADC, DAC, FPGA (Xilinx Spartan 3-500e) and differential LVDS line drivers for serial high-speed links, as shown in Fig. 16. It is used primarily here for comparison with the software architecture.

Hardware implementations represent a number of advancements. For example, the hardware implementation can split the double word access into two memory cycles or use 32-bit memory for stacks, always providing one-cycle memory access. For performance reasons, the hardware implementation can implement stacks with block RAM components and single registers holding the first top values, enabling parallel computations on stack elements.

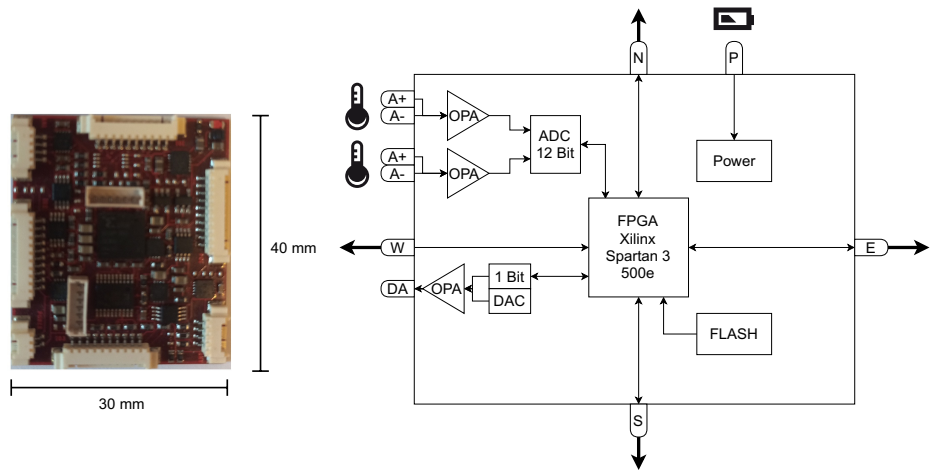

**Fig. 16.** FPGA sensor node with wired communication (Xilinx SRAM FPGA Spartan 3-500e)

## 7.3  Pocket GUW Laboratory

The pocket GUW laboratory, as shown in Fig. 17, is a low-cost measuring device systeme close to the sensor node introduced and aims to support software and algorithm development for damage diagnostics using guided ultrasonic waves (GUW): The software framework consists of two separated applications communicating via the system call-gate interface (see Sec. 3.7): base oscilloscope application with ADC and DAC device drivers (32 kB ROM, 24 kB RAM), and the REXA-VM (10 kB ROM, 20 kB RAM).

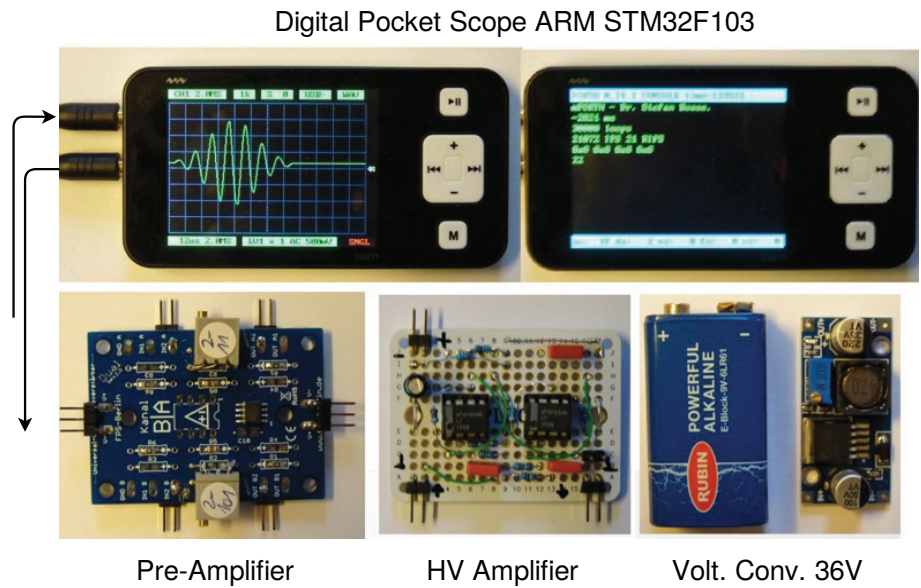

**Fig. 17.** The pocket GUW laboratory only using low-budget and low-quality devices for GUW-based damage detection in Fibre composite materials. The DSO implements REXA-VM and communicates via an USB virtCOM port with an external computer.

The main osclllloscope application initializes the VM and attachs callback functions and all IOS functions and data structures. Code is received by the main application and send to the VM compiler. The code segment is shared by both applications. Console output and VM communication is passed to callback functions installed by the main application. The IAR Embedded Workbench for ARM compiler used in this work, which generates high-quality, optimized machine code but caps the code size at 32 kB for the non-profit free license version, is the primary driver behind this two-program strategy. There shouldn't be any license fees for the compiler software in order to make it available to a large community, including students and scholars.

### 7.4 Simple Measuring Job

A simple measuring job that is controlled by a remote master program and sent to the GUW laboratory device enables a signal generator, starts a signal acquisition, and sends the sampled sensor data to the master node, as shown in Ex. 3. The signal processing operations *adc* and *dac* are provided by the sensor node host application. The sensor node is connected via a serial link to a remote computer.

**Ex. 3.** Simple measuring job program template used for active GUW signal sampling starting a signal generator providing a hamming window sine-wave burst signal (dac) and the signal sampling (adc). After the sampling task is completed, the data is transferred to the remote controller via a serial USB link (loop).

```
 1: ( A full measuring job as a FORTH program with embedded data )
 2: <waveTable> <intervalMS>  <dacDiv>    <sampleFreq> dac
 3: <syncMode>   <sampleDepth> <gainMode> <sampleFreq> adc
 4: 300 1 sampled await
 5: var offset sample0 read offset !
 6: <samples> 0 do
 7:    offset @ samples read
 8:    out ( send data value )
 9:    offset @ 1 + <sampleDepth> mod offset !
10: loop
11: end
```

### 7.5 Structural Health Monitoring

The GUW pocket laboratory was used in addition to simple and low-cost signal electronics for damage diagnostics with a fiber-metal laminate (FML) plate. The plate was equipped with integrated piezoelectric transducers. The set-up consists of:

- Pair of transducers;

- A 100-line code REXA VM program generates and acquires signals for active measurement.

- Hull computation using a rectifier-low-pass algorithm to approximate an analytical signal;

- A simple ANN that predicts damage and distance;

- On the plate's surface, moveable pseudo defects (neodymium magnets) simulate an internal defect.

The DSO was connected to a computer running a node-webkit (nw.js) applciation and directly accessing the REXA VM. The application software performs measuring jobs by sending code to the REXA VM and waiting for the measuring results. The results were stored in a SQL database for later processing and analysis. The framework using the REXA VM creates a compact measuring laboratory.

## 8. Results and Evaluation

### 8.1 Ressources

The main advantage of the proposed VM architecture is the capability to create the main and crucial parts of the VM using code generators and adapt the VM architecture to specific applications and host architectures.

All data (and code) memory is allocated statically at compile time. There is no dynamic memory management requirement. The compiler works in-place, i.e., it compiles source text stored in free regions of the CS directly in-place into bytecode (no additional storage space is required during compilation).

Although the data, return, and loop stacks DS, RS, and FS, respectively, can be small, multi-threading and multi-tasking increase the stack storage requirements by the number of maximally supported threads and scheduled tasks.

| Target | Configuration | ROM | RAM |
|---|---|---|---|
| STM32 F103C3, 72MHz, 256kB ROM, 48kB RAM | CS=1024, DS=256, RS=128, FS=64, Words=101 | 8.2kB | 8.1kB |
| STM32 F103C3, 72MHz, 256kB ROM, 48kB RAM | CS=1024, DS=256, RS=128, FS=64, Words=64 (no double word arithmetic) | 7.1kB | 7.5kB |
| STM32 L031, 16MHz, 32kB ROM, 8kB RAM | CS=1024, DS=256, RS=32, FS=32, Words=64 (no double word arithmetic) | 7.1kB | 8kB |
| STM32 F103C3, 72MHz, 256kB ROM, 48kB RAM | CS=4096, DS=1024, RS=256, FS=128, Words=101 | 8.2kB | 15kB |
| i5-7300U, 2GHz | CS=16384, DS=4096, RS=1024, FS=256, Words=101 | 32kB | 64kB |

**Tab. 7.** Summary of software VM resource requirements for various configurations and different target platforms. The performance giving the number of Forth word instructions per second is measured with a simple calibrated benchmark.

The hardware resources required for typical REXA VM configurations can be found in Tab. 8. Two different FPGA technologies were compared: SRAM-based and EEPROM flash-cells for ROM and configuration implementations. The results were retrieved by the Xilinx ISE (11) and Synplicity FPGA compiler (89).

| Target | Configuration | Clock [MHz] | Slices | DFF | Block RAM |
|--------|---------------|-------------|--------|-----|-----------|
| XC3S500e SRAM FPGA | CS=4096, DS=1024, SS/RS=32, Words=84 | 70 | 2029/4565 | 700/9312 | 9/20 |

**Tab. 8.** Summary of software VM resource requirements for various configurations and different target platforms. The performance giving the number of Forth word instructions per second is measured with a simple calibrated benchmark.

## 8.2 Performance

**MWPS(MCPS)**

The performance of the REXA VM on various host architectures and processors is measured in million bytecode instruction word executions per second (MWPS) and the compiler performance in million word compilations per second (MCPS), as shown in Tab. 9. The efficiency ε vm is now proportional to MWPS, and the memory is proportional to the CS/DS/RS/FS segments.

| Target | Configuration | MIPS | MCPS | $\varepsilon_{vm}$ |
|--------|---------------|------|------|--------------------|
| STM32 F103C3, 72MHz, 256kB ROM, 48kB RAM | CS=1024, DS=256, RS=128, FS=64, Words=101 | 1.1 | 0.1 | 6.6 |
| STM32 F103C3, 72MHz, 256kB ROM, 48kB RAM | CS=1024, DS=256, RS=128, FS=64, Words=64 (no double word arithmetic) | 1.1 | 0.1 | $6.6^{-3}$ |
| STM32 F103C3, 72MHz, 256kB ROM, 48kB RAM | CS=4096, DS=1024, RS=256, FS=128, Words=101 | 1.1 | 0.1 | $13.2^{-3}$ |
| STM32 L031, 16 MHz, 32 kB ROM, 8 kB RAM | CS=1024, DS=256, RS=32, FS=32, Words=101 | 0.24 | 0.02 | 0.48 |
| i5-7300U, 2GHz (A=400 mm², P=10W) | CS=16384, DS=4096, RS=1024, FS=256, Words=101 | 280 | 27 | $1.4^{-3}$ |

**Tab. 9.** Summary of VM performance giving the number of executed Forth word instructions and the number of word compilations per second measured with a simple calibrated benchmark.

**DSP/ANN**

The DSP/ANN module of REXA VM was tested with different ANN configurations, with 3-5 layers, and 2-64 neurons per layer. The measured results for the ARM Cortex M0, STM32 L031 and F103 microcontrollers are shown in Fig. 18 and Tab. 10 (larger networks can only be handled by the F103 due to RAM allocation). The code size includes the ANN parameter data, input and output vectors, and the forward computation function. The code size ranges from 200 to 6K bytes for 6-100 neurons. As shown in Fig. 18, the computation time for one neuron decreases with increasing network size. This is a result of the VM execution overhead that dominates for very small network sizes. The average computation time is about 700 micros / neuron / MHz for the ARM Cortex STM32 microcontrollers. Note that the ARM processors underperform compared with modern Intel x86/x64 processors (but still have an efficiency ratio of 1:100 if power and chip area are considered). Typical computation times for medium sized networks are in the millisecond range, fully suitable for on-node classification.

| Layers | Neurons | Code [Bytes] | Forward Time [ms/MHz] |
|---|---|---|---|
| [2,3,1] | 6 | 237 | 7.7 |
| [4,3,2] | 9 | 281 | 8.2 |
| [4,6,2] | 12 | 336 | 9.1 |
| [4,8,2] | 14 | 372 | 11.2 |
| [4,8,4] | 16 | 416 | 10.5 |
| [4,8,8,2] | 22 | 601 | 14.4 |
| [4,8,8,4] | 24 | 645 | 16.6 |
| [4,8,8,8,4] | 32 | 874 | 21.0 |
| [4,32,2] | 38 | 804 | 17.1 |
| [8,32,32,8] | 80 | 3813 | 43.5 |
| [8,64,32,8] | 112 | 6566 | 58.3 |

**Tab. 10.** Normalized ANN results on STM32 platform

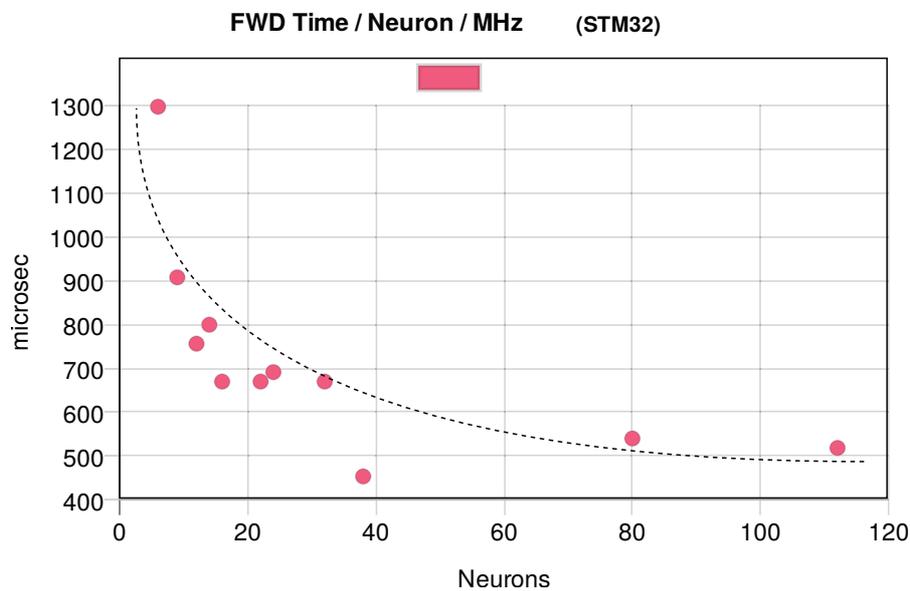

**Fig. 18.** ANN forward computation times for one neuron and per MHZ clock frequency (different ANN architectures with 3-5 layers and 2-64 neurons per layer)

## 9. Conclusions

A stack-based virtual machine architecture for low-resource, tiny embedded systems was introduced and analyzed. The overhead, even on very low-resource systems, is low with respect to typical running times under energy constraints and tasks to be performed in real-time (i.e., under time constraints). The unified VM architecture can be implemented in both software and hardware. A major feature is the tight bundling of a text-to-bytecode compiler with the bytecode interpreter, ensuring robustness, security, stability, and interoperability in strong heterogenous environments. The software and hardware implementations are equivalent at the implementation level. A strong focus was laid on resource sharing, e.g., an AD conversion sample buffer can be used for signal processing computations directly from the VM programming level. Even on very low-resource embedded systems, machine learning models can be implemented and computed. The model parameters and configuration, e.g., for an artificial neural network, are embedded in the program code. There is no data heap. Program data is stored in the code segment, organized in code frames.

The VM is highly customizable and extensible. Parts of the VM program code are created by parametrizable code generators. The ISA can be extended by the IOS, too, providing programming level access to host application functions and data, e.g., for accessing ADC and DAC devices.